\documentclass[conference,a4paper]{IEEEtran}

\usepackage[utf8]{inputenc} 
\usepackage[T1]{fontenc}
\usepackage{url}
\usepackage{hyperref}
\usepackage{ifthen}
\usepackage{cite}
\usepackage{stfloats}
\usepackage{lipsum}
\usepackage{cuted}
\usepackage[cmex10]{amsmath} 
\usepackage[pdftex]{graphicx}
\ifCLASSOPTIONcompsoc
  \usepackage[caption=false,font=normalsize,labelfont=sf,textfont=sf]{subfig}
\else
\usepackage[caption=false,font=footnotesize]{subfig}
\fi
\usepackage{enumitem}
\DeclareGraphicsExtensions{.pdf}

\usepackage{amssymb}
\newtheorem{theorem}{Theorem}

\newtheorem{remark}{Remark}
\newcounter{MYtempeqncnt}

\begin{document}
\title{The Secure Two-Receiver Broadcast Channel With One-Sided Receiver Side Information} 

\author{%
   \IEEEauthorblockN{Jin Yeong Tan, Lawrence Ong, and Behzad Asadi}
   \IEEEauthorblockA{School of Electrical Engineering and Computing, The University of Newcastle, Newcastle, Australia\\
                     Email: jinyeong.tan@uon.edu.au, lawrence.ong@newcastle.edu.au, behzad.asadi@uon.edu.au}
 }

\maketitle

\begin{abstract}
	This paper studies the problem of secure communication over the two-receiver discrete memoryless broadcast channel with one-sided receiver side information and with a passive eavesdropper. We proposed a coding scheme which is based upon the superposition-Marton framework. Secrecy techniques such as the one-time pad, Carleial-Hellman secrecy coding and Wyner secrecy coding are applied to ensure individual secrecy. This scheme is shown to be capacity achieving for some cases of the degraded broadcast channel. We also notice that one-sided receiver side information provides the advantage of rate region improvement, in particular when it is available at the weaker legitimate receiver.
\end{abstract}



\section{Introduction}

\subsection{Background}

Broadcast channels model typical downlinks in communication systems, where a transmitter broadcasts independent messages to multiple receivers.
Similar to other wireless communication channels, the open nature of broadcast channels make them susceptible to passive eavesdroppers present in the communication range. Due to the widespread application of broadcast channels in communication systems, it is crucial to ensure secure communication over broadcast channels.

The problem of secure broadcast channel has been studied by Csisz$\acute{\text{a}}$r and K$\ddot{\text{o}}$rner\cite{iref1}, Chia and El Gamal\cite{iref2} as well as Schaefer and Boche \cite{Schaefer_Boche14}. These works \cite{iref1,iref2,Schaefer_Boche14} studied cases of two- or three-receiver broadcast channel in which a common message is transmitted to all receivers and a private message is protected from a certain number of receivers. Similarly, Chen, Koyluoglu and Sezgin investigated the problem of secure two-receiver broadcast channel but it involves the protection of two private messages from an eavesdropper \cite{cref1}. 

The works discussed above were also extended to account for the availability of complementary receiver side information at the legitimate receivers which is unknown to the eavesdropper. The availability of complementary receiver side information implies that each receiver knows a priori the message they need not decode and it is shown to help in improving the secrecy rate region \cite{iref13,iref14,cref2}.

However, when the available receiver side information is non-complementary, i.e., receiver side information is missing on certain number of receivers, it is generally unknown if the advantage of secrecy rate region improvement still holds. This motivates us to study the effect of non-complementary receiver side information on secure broadcast channel by considering the two-receiver discrete memoryless broadcast channel with one-sided receiver side information and with an eavesdropper. We will also be considering the individual secrecy notion which requires the individual information leakage from each message to the eavesdropper to be vanishing \cite{iref14,iref15,cref1,cref2}. In short, this paper aims to gain insights on the advantages of non-complementary receiver side information on secure broadcasting by proposing a secrecy scheme and deriving the individual secrecy rate region for the two-receiver discrete memoryless broadcast channel with one-sided receiver side information and with a passive eavesdropper. 

\subsection{Contributions}

Existing works on secure broadcast channel consider only symmetrical receiver side information setup, i.e., receiver side information is either absent \cite{iref1,iref2,Schaefer_Boche14,cref1} or present at both receivers \cite{iref13,iref14,cref2,Me1}.  The usage of the coding schemes in these works for the non-complementary receiver side information setting in this paper poses some limitations. More precisely, schemes for the former case are suboptimal since they ignore the availabity of receiver side information, whereas schemes for the latter case cannot be applied if receiver side information is absent at at least one receiver. As a result, we propose a technique to deal with this asymmetrical receiver side information setting. Using the proposed coding scheme, we derive a general inner bound for the two-receiver discrete memoryless broadcast channel with one-sided receiver side information and with a passive eavesdropper under individual secrecy constraints. We show that this inner bound establishes the individual secrecy capacity region for the physically degraded deterministic broadcast channel under all permutations of channel degradedness. Through the capacity region results, we observe that the advantage of one-sided receiver side information manifests itself when it is available at the statistically weaker receiver, providing gains in the capacity rate region.

\subsection{Paper Organization}

The entire paper will be organized as follows. Section II will focus on the system model. Section III will provide the main results on the coding scheme simplification. Next, section IV will present some capacity region results. Lastly, Section V will conclude the paper. 

\section{System Model}

In this paper, we will denote random variables by uppercase letters, their corresponding realizations by lowercase letters and their corresponding sets by calligraphic letters. A $(j-i+1)$-sequence of random variables will be denoted by $X_i^j=(X_i,…,X_j)$ for $1\leq i\leq j$. Whenever $i=1$, the subscript will be dropped, resulting in $X^j=(X_1,…,X_j)$. $\mathbb{R}^d$ represents the $d$-dimensional real Euclidean space and $\mathbb{R}_+^d$ represents the $d$-dimensional non-negative real Euclidean space. $\mathcal{R}$ will be used to represent a subset of $\mathbb{R}^d$. $\mathcal{T}^n_{\epsilon}$ represents the set of jointly $\epsilon$-typical $n$-sequences. Meanwhile, $[a:b]$ refers to a set of natural numbers between and including $a$ and $b$, for $a\leq b$. Lastly, the operator $\times$ denotes the Cartesian product.

\begin{figure}[t]
	\centering
	\includegraphics[scale=0.55]{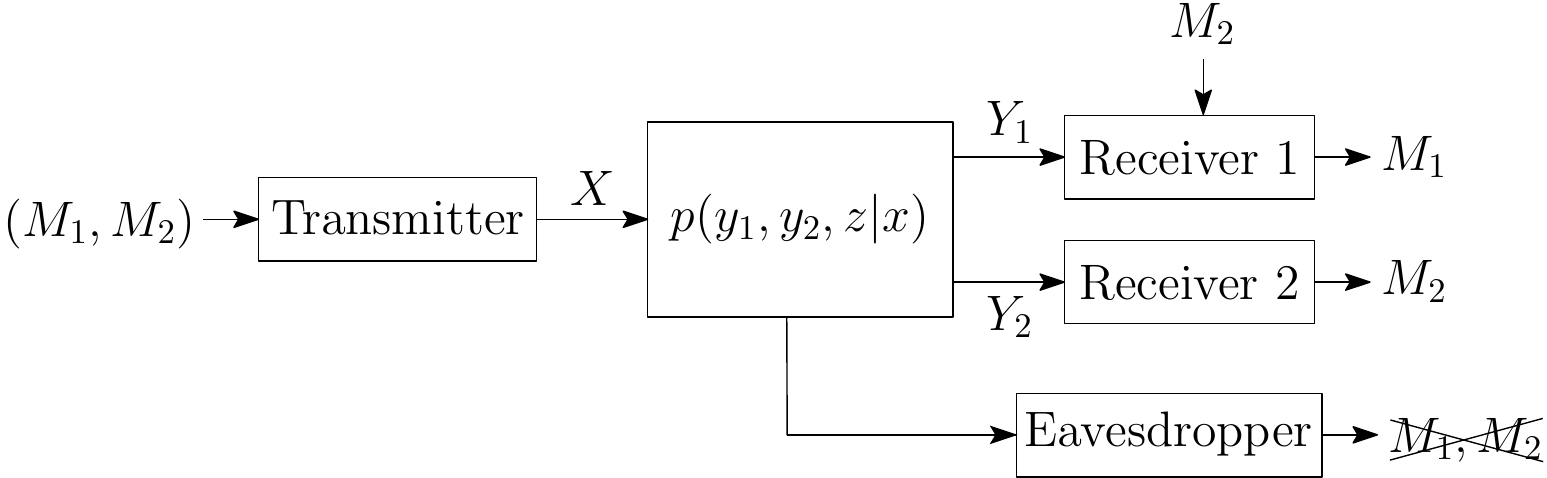}
	\caption{The two-receiver broadcast channel with one-sided receiver side information in the presence of an eavesdropper.}
	\label{fig:1}
\end{figure}
The paper focuses on devising a coding scheme for the two-receiver discrete memoryless broadcast channel with one-sided receiver side information and with a passive eavesdropper (hereafter referred as the broadcast channel with one-sided receiver side information). The system model for this case is illustrated in Fig.~\ref{fig:1}. In this model, we define $(M_1, M_2)$ as the source messages, $M_i$ as the message requested by legitimate receiver $i$, for all $i\in\{1,2\}$. Let $X$ denote the channel input from the sender, while $Y_i$ and $Z$ denote the channel output to receiver $i$ and the eavesdropper respectively. In $n$ channel uses,  $X^n$ represents the transmitted codeword, $Y_i^n$ represents the signal received by legitimate receiver $i$ and $Z^n$ represents the signal received by the eavesdropper. The memoryless (and without feedback) nature of the channel also implies that
\begin{align}
&p(y_1^n,y_2^n,z^n|x^n)=\prod_{i=1}^n p(y_{1i},y_{2i},z_i|x_i).
\end{align} 

\begin{figure*}[b]
	\centerline{\subfloat[Rate splitting]{\includegraphics[scale=0.5]{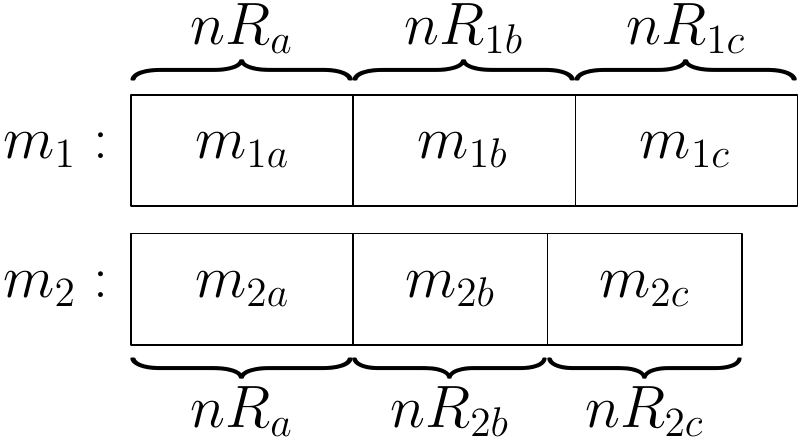}
			\label{fig:2a}}
		\hfil
		\subfloat[Encoding]{\includegraphics[scale=0.6]{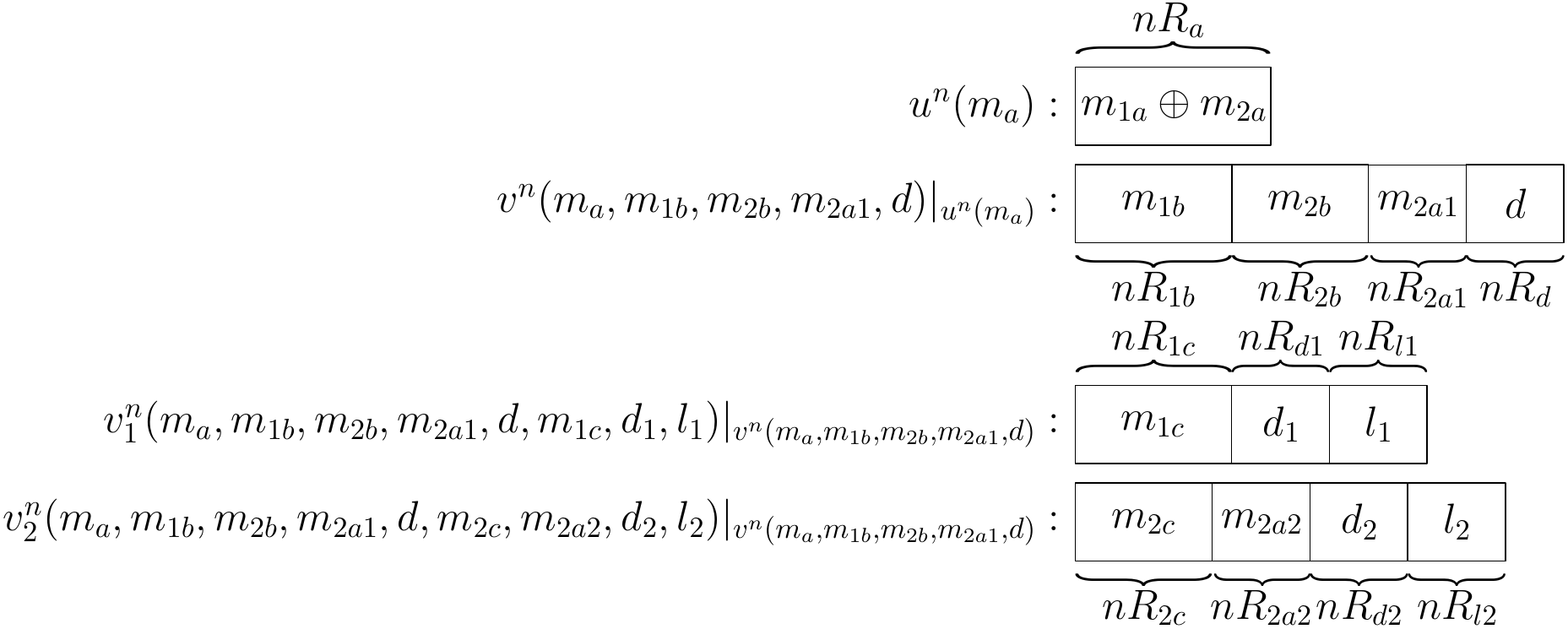}
			\label{fig:2b}}}
	\caption{Proposed coding scheme for the two-receiver broadcast channel with one-sided receiver side information and with an eavesdropper.}
	\label{fig:2}
\end{figure*}

In this case, the transmitter will be sending messages $M_{1}$ and $M_{2}$ to legitimate receiver 1 and 2, respectively through the channel $p(y_{1},y_{2},z|x)$. Besides, only one of the legitimate receivers will be having the message they do not need to decode as receiver side information to aid them in decoding the transmitted messages. For our discussion, we will work with the case in which receiver 1 knows $M_2$ a priori.

\textit{Definition 1:} A $(2^{nR_1},2^{nR_2},n)$ secrecy code for the two-receiver discrete memoryless broadcast channel with one-sided receiver side information consists of:
\begin{itemize}
	\item two message sets, where $\mathcal{M}_1=[1:2^{nR_1}]$ and $\mathcal{M}_2=[1:2^{nR_2}]$;
	\item an encoding function, $f:\mathcal{M}_1\times\mathcal{M}_2\rightarrow\mathcal{X}^n$, such that $X^n=f(M_1,M_2)$; and
	\item two decoding functions, where $g_1:\mathcal{Y}_1^n\times\mathcal{M}_{2}\rightarrow\mathcal{M}_1$, such that $\hat{M}_1=g_1(Y_1^n, M_2)$ at receiver 1 and $g_2:\mathcal{Y}_2^n\rightarrow\mathcal{M}_2$, such that $\hat{M}_2=g_2(Y_2^n)$ at receiver 2.
\end{itemize}

Both messages, $M_1$ and $M_2$ are assumed to be uniformly distributed over their respective message set. Hence, we have $R_i=\frac{1}{n}H(M_i)$, for all $i\in\{1,2\}$. Meanwhile the individual information leakage rate associated with the $(2^{nR_1},2^{nR_2},n)$ secrecy code is defined as $R_{\text{L},i}^{(n)}=\frac{1}{n}I(M_i;Z^n)$, for all $i\in\{1,2\}$.

The probability of error for the secrecy code at each receiver $i$ is defined as $P_{\text{e},i}^{(n)}=P\{\hat{M}_i\ne M_i\}$, for $i\in\{1,2\}$. A rate pair $(R_1,R_2)$ is said to be achievable if there exists a sequence of $(2^{nR_1},2^{nR_2},n)$ codes such that 
\begin{align}
&P_{\text{e},i}^{(n)}\leq \epsilon_n\text{, for all }i\in\{1,2\}\\			
&R_{\text{L},i}^{(n)}\leq \tau_n\text{, for all }i\in\{1,2\}\label{sec}\\		
& \lim\limits_{n\rightarrow\infty}\epsilon_n=0\text{ and }\lim\limits_{n\rightarrow\infty}\tau_n=0 			
\end{align}

\section{Coding Scheme for Broadcast Channel With One-Sided Receiver Side Information}

Our coding scheme is constructed by having the superposition-Marton coding scheme \cite{cref6},\cite{cref5} as a basic framework. The message segments encoded at each layer of the scheme is then protected using secrecy techniques such as the one-time pad \cite{cref8}, Carleial-Hellman secrecy coding \cite{cref3} and Wyner serecy coding \cite{cref4}.  

As illustrated in Fig.~\ref{fig:2a}, this coding scheme involves splitting $M_i$, for all $i\in\{1,2\}$, into three independent message segments, namely $M_{ia}$ at rate $R_{ia}$, $M_{ib}$ at rate $R_{ib}$ and $M_{ic}$ at rate $R_{ic}$, i.e., $M_i=(M_{ia},M_{ib},M_{ic})$ where $M_{ia}\in [1:2^{nR_{ia}}]$, $M_{ib}\in [1:2^{nR_{ib}}]$, $M_{ic}\in [1:2^{nR_{ic}}]$ and $R_i=R_{ia}+R_{ib}+R_{ic}$. 

Exploiting the availability of receiver side information at receiver 1, we will be XOR-ing the first segment of each message to form a one-time pad signal which ensures individual secrecy. The one-time pad signal $M_a$ has a rate of $R_a$ and $M_a=M_{1a}\oplus M_{2a}$ where $M_{a}\in [1:2^{nR_{a}}]$. This XOR-ing process gives us another rate relation $R_a=R_{1a}=R_{2a}$. Note that since receiver 2 does not have any receiver side information, it is unable to retrieve $M_{2a}$ even though $M_a$ is successfully decoded. As a result, we introduce a new approach of further splitting $M_{2a}$ into $M_{2a1}$ at rate $R_{2a1}$ and $M_{2a2}$ at rate $R_{2a2}$, i.e., $M_{2a}=(M_{2a1},M_{2a2})$ where $M_{2a1}\in [1:2^{nR_{2a1}}]$, $M_{2a2}\in [1:2^{nR_{2a2}}]$ and $R_{2a}=R_{2a1}+R_{2a2}$. $M_{2a1}$ and $M_{2a2}$ will be transmitted to receiver 2 through different codeword layers, allowing $M_{2a}$ to be fully recovered.

These message segments then make up the coding scheme in Fig.~\ref{fig:2b} which comprises of four layers: cloud center codeword $U^n$, common satellite codeword $V^n$ as well as private satellite codewords $V_1^n$ and $V_2^n$. The cloud center codeword $U^n$ holds the one-time pad signal $M_{1a}\oplus M_{2a}$, i.e., $M_a$. Upon decoding the one-time pad signal, receiver 1 can unveil the intended message segment $M_{1a}$ using its receiver side information $M_2$. 

Next, a common satellite codeword $V^n$ is formed and combined with $U^n$ using superposition coding \cite{cref6}. In the $V^n$ codeword layer, Carleial-Hellman secrecy coding \cite{cref3} and Wyner serecy coding \cite{cref4} is implemented simultaneously to protect the message segments $M_{1b}$, $M_{2b}$ and $M_{2a1}$. The implementation of Wyner serecy coding is apparent since a randomly generated $D$ is used to protect $M_{1b}$, $M_{2b}$ and $M_{2a1}$. Carleial-Hellman secrecy coding is applied since it allows each of the message segments $M_{1b}$ and $M_{2b}$ to act as additional randomness in protecting each another and $M_{2a1}$. This helps reduce the size of randomness $D$ and subsequently increases the size of $M_{1b}$ and $M_{2b}$ which bear useful information. 

The private satellite codewords $V_1^n$ and $V_2^n$ are responsible of carrying the remaining message segments to their respective legitimate receivers. In our case, $V_1^n$ carries $M_{1c}$ to receiver 1, whereas $V_2^n$ carries $M_{2a2}$ and $M_{2c}$ to receiver 2. Both $V_1^n$ and $V_2^n$ are also secured with Wyner secrecy coding \cite{cref4} using randomly generated $D_1$ and $D_2$ respectively. The private satellite codewords $V_1^n$ and $V_2^n$ are first combined under Marton coding in which the components $L_1$ and $L_2$ are present to ensure the existence of at least one jointly typical sequence pair between $V_1^n$ and $V_2^n$ \cite{cref5}. Note that $L_1$ and $L_2$ do not play a role in message protection since they cannot be preselected \cite{Mansour_Schaefer_Boche16}. The $V_1^n$ and $V_2^n$ codeword layers are then combined with the upper $U^n$ and $V^n$ codeword layers using superposition coding, completing the construction of our coding scheme. This coding scheme achieves the individual secrecy rate region $\mathcal{R}$ in Theorem~\ref{theorem1}. 

\begin{theorem}\label{theorem1}
	The following individual secrecy rate region is achievable for the two-receiver discrete memoryless broadcast channel with one-sided receiver side information and with a passive eavesdropper:
	\begin{equation}\label{eq2}
	\mathcal{R} \triangleq
	\left\{
	\begin{split}
	(R_1,R_2)\\ \in \mathbb{R}_+^2
	\end{split}
	\left\vert
	\begin{split}
	&R_1<I(V_1;Y_1|V)-I(V_1;Z|V)\\
	&+I(V;Y_2|U)-I(V;Z|U)+R_{N1}+R_{N2},\\
	&R_1<I(U,V,V_1;Y1)-I(V_1;Z|V)+R_{N3},\\
	&R_1-R_2<I(V,V_1;Y_1|U)\\&-I(V,V_1;Z|U)+R_{N3},\\
	&R_2<I(V,V_2;Y_2|U)-I(V,V_2;Z|U)\\
	&+\text{min}\{R_{N2}+R_{N4},0\},\\
	&R_1+R_2<I(U,V,V_2;Y_2)\\&-I(V,V_2;Z|U)
	+R_{N4}+R_{N5}\\
	&\text{over all }p(u)p(v|u)p(v_1,v_2|v)p(x|v_1,v_2)\\
	&\text{subject to }\\
	&I(V,V_1;Y_1|U)+R_{N3}>I(V,V_1;Z|U),\\
	&I(V_1;Y_1|V)+R_{N3}>I(V_1;Z|V),\text{ and }\\
	&I(V_2;Y_2|V)>I(V_2;Z|V)\text{.}\\ 
	\end{split}
	\right.
	\right\}
	\end{equation}
\end {theorem}
where $R_{N1}=I(V_2;Y_2|V)-I(V_2;Z|V)-I(V_1;V_2|V)$, $R_{N2}=\text{min}\{I(V;Y_1|U)-I(V;Z|U),0\}$, $R_{N3}=\text{min}\{R_{N1},0\}$, $R_{N4}=I(V_1;Y_1|V)-I(V_1;Z|V)-I(V_1;V_2|V)$ and $R_{N5}=\text{min}\{I(V;Y_1|U)+I(U;Y_1)-I(U;Y_2), I(V;Y_1|U), I(V;Z|U)\}$.

\begin{IEEEproof}[Proof of Theorem~\ref{theorem1}] Refer to Appendix~\ref{AppB} for the complete achievability proof.
\end{IEEEproof}

\section{Capacity Region in Some Special Cases}
In this section, we will show that our scheme is optimal in achieving the individual secrecy capacity region for some special cases of the physically degraded discrete memoryless broadcast channel defined as 
$p(y_1,y_2,\dotsc,y_k|x)=p(y_1|x)p(y_2|y_1)\dotsb p(y_k|y_{k-1})$, i.e., $X \rightarrow Y_1 \rightarrow Y_2 \rightarrow \dotsc \rightarrow Y_k$ forms a Markov chain.

More specifically, the individual secrecy capacity region for the physically degraded deterministic broadcast channel under all permutations of channel degradedness among $Y_1$, $Y_2$ and $Z$ is proven in the following theorems. In Theorem~\ref{theorem2}, the receiver with receiver side information is the weaker one, whereas in Theorem~\ref{theorem3}, the receiver with receiver side information is the stronger one.

\begin{theorem}\label{theorem2}
	For the two-receiver discrete memoryless broadcast channel with one-sided receiver side information and with a passive eavesdropper such that
	\begin{enumerate}
		\item $Y_1$, $Y_2$ and $Z$ are deterministic functions of $X$; and
		\item $X \rightarrow Y_2 \rightarrow Y_1 \rightarrow Z$, or \newline 
		$X \rightarrow Y_2 \rightarrow Z \rightarrow Y_1$, or \newline
		$X \rightarrow Z \rightarrow Y_2 \rightarrow Y_1$, 
	\end{enumerate}
	the individual secrecy capacity region is given by 
	\begin{equation}\label{eq3}
	\mathcal{R}_1 \triangleq
	\left\{
	\begin{split}
	(R_1,R_2)\\ \in \mathbb{R}_+^2
	\end{split}
	\left\vert
	\begin{split}
	&R_1\leq \text{min}\{H(Y_1),H(Y_2|Z),\\
	&\qquad\quad H(Y_1|Z)+R_2\},\\
	&R_2\leq H(Y_2|Z),\\
	&R_1+R_2\leq H(Y_2)\\
	&\text{over all } p(x).
	\end{split}
	\right.
	\right\}.
	\end{equation}
\end {theorem}

\begin{IEEEproof}[Proof of Theorem~\ref{theorem2}] We present the achievability proof in the following, and the converse proof in Appendix~\ref{AppB}.
The region $\mathcal{R}_1$ can be achieved by taking the union of the regions: $\mathcal{R}_{1a}$, $\mathcal{R}_{1b}$, $\mathcal{R}_{1c}$ and $\mathcal{R}_{1d}$ stated below.
\begin{enumerate}[leftmargin=*]
	\item For all $p(x)$ such that $H(Y_2)>H(Y_1)>H(Z)$\newline
	The individual secrecy rate region in this case can be achieved using Theorem~\ref{theorem1} by setting $U=\phi$, $V_1=V=Y_1$ and $V_2=Y_2$. This reduces the proposed scheme to a two-layer superposition coding scheme in which $V^n$ and $V_2^n$ need to be decoded. As a result, from the proposed coding scheme in Fig.~\ref{fig:2b}, we remove $M_{a}$, $M_{1a}$, $M_{2a}$, $M_{2a1}$, $M_{2a2}$, $M_{1c}$, $D_{1}$, $L_{1}$ and $L_{2}$. Upon removing the corresponding rate terms and constraints, we apply the Fourier-Motzkin procedure to obtain the region $\mathcal{R}_{1a}$ as follows:
	\begin{equation}\label{eq4}
	\left\{
	\begin{split}
	(R_1,R_2)\\ \in \mathbb{R}_+^2
	\end{split}
	\left\vert
	\begin{split}
	&R_1< \text{min}\{H(Y_1),H(Y_2|Z),\\
	&\qquad\quad H(Y_1|Z)+R_2\},\\
	&R_2< H(Y_2|Z),\\
	&R_1+R_2< H(Y_2)\\
	&\text{over all }p(x)\text{ subject to }\\
	&H(Y_1|Z)>0\text{ and } H(Y_2|Y_1)>0.\\
	\end{split}
	\right.
	\right\}.
	\end{equation}
	\item For all $p(x)$ such that $H(Y_2)=H(Y_1)>H(Z)$\newline
	In order to achieve the individual secrecy rate region in this case, from Theorem~\ref{theorem1}, we set $U=\phi$ and $V_1=V_2=V=Y_1=Y_2$. This reduces the proposed scheme to a single $V^n$ codeword layer. Thus, from the proposed coding scheme in Fig.~\ref{fig:2b}, we are left with $M_{1b}$, $M_{2b}$ and $D$. Upon removing the corresponding rate terms and constraints, we apply the Fourier-Motzkin procedure to obtain the region $\mathcal{R}_{1b}$ as follows:
	\begin{equation}\label{eq5}
	\left\{
	\begin{split}
	(R_1,R_2)\\ \in \mathbb{R}_+^2
	\end{split}
	\left\vert
	\begin{split}
	&R_1< H(Y_2|Z),\\
	&R_2< H(Y_2|Z),\\
	&R_1+R_2< H(Y_2)\\
	&\text{over all }p(x)\text{ subject to }\\
	&H(Y_1|Z)>0.\\
	\end{split}
	\right.
	\right\}.
	\end{equation}
	\item For all $p(x)$ such that $H(Y_2)> H(Z)\geq H(Y_1)$\newline
	When the eavesdropper is of equal channel strength as the weaker legitimate receiver 1, from Theorem~\ref{theorem1}, we need to set $V_1=V=U=Y_1$ and $V_2=Y_2$. This reduces the proposed scheme to a two-layer superposition coding scheme with the $U^n$ and $V_2^n$ codewords. Thus, from the proposed coding scheme in Fig.~\ref{fig:2b}, we remove $M_{1b}$, $M_{2a1}$, $M_{2b}$, $M_{1c}$, $D$, $D_{1}$, $L_{1}$ and $L_{2}$. Upon removing the corresponding rate terms and constraints, we apply the Fourier-Motzkin procedure to obtain the region $\mathcal{R}_{1c}$ as follows:
	\begin{equation}\label{eq6}
	\left\{
	\begin{split}
	(R_1,R_2)\\ \in \mathbb{R}_+^2
	\end{split}
	\left\vert
	\begin{split}
	&R_1< \text{min}\{H(Y_1),R_2\},\\
	&R_2< H(Y_2|Z),\\
	&R_1+R_2< H(Y_2)\\
	&\text{over all }p(x).
	\end{split}
	\right.
	\right\}.
	\end{equation}
	\item For all $p(x)$ such that $H(Z)\geq H(Y_2)\geq H(Y_1)$\newline
	When the eavesdropper is not weaker than both legitimate receivers, from Theorem~\ref{theorem1}, we need to set $V_1=V_2=V=U=\phi$. This essentially means that secure communication is not possible and we obtain the region $\mathcal{R}_{1d}=\{(0,0)\}$.
\end{enumerate}
By taking the union of $\mathcal{R}_{1a}$, $\mathcal{R}_{1b}$, $\mathcal{R}_{1c}$ and $\mathcal{R}_{1d}$, we achieve the individual secrecy capacity region~\eqref{eq3}. 
\end {IEEEproof}

\begin{remark}
Referring to the individual secrecy capacity region for the two-receiver discrete memoryless broadcast channel without receiver side information and with a passive eavesdropper \cite[Theorem 3]{cref1}, we flip the order of channel degradedness between $Y_1$ and $Y_2$, and set $Y_1$ to be a deterministic function of $X$. Comparing that with our results in Theorem~\ref{theorem2} shows that the availability of receiver side information at receiver 1, in particular, the weaker legitimate receiver, allows us to achieve a larger individual secrecy rate region.
\end{remark}
\begin{remark}
It is also interesting to note that while using the superposition-Marton coding scheme \cite{cref6},\cite{cref5}, the inclusion of one-time pad signal, a network coding solution, can help establish a secure communication when the eavesdropper is stronger than the weaker legitimate receiver. This can be observed from the achievability proof of $\mathcal{R}_{1c}$. However, when the eavesdropper is not weaker than both legitimate receivers, secure communication is not possible.
\end{remark}

\begin{theorem}\label{theorem3}
	For the two-receiver discrete memoryless broadcast channel with one-sided receiver side information and with a passive eavesdropper such that
	\begin{enumerate}
		\item $Y_1$, $Y_2$ and $Z$ are deterministic functions of $X$; and
		\item $X \rightarrow Y_1 \rightarrow Y_2 \rightarrow Z$, or \newline 
		$X \rightarrow Y_1 \rightarrow Z \rightarrow Y_2$, or \newline
		$X \rightarrow Z \rightarrow Y_1 \rightarrow Y_2$, 
	\end{enumerate}
	the individual secrecy capacity region is given by 
	\begin{equation}\label{eq8}
	\mathcal{R}_2 \triangleq
	\left\{
	\begin{split}
	(R_1,R_2)\\ \in \mathbb{R}_+^2
	\end{split}
	\left\vert
	\begin{split}
	&R_1\leq H(Y_1|Z),\\
	&R_2\leq H(Y_2|Z),\\
	&R_1+R_2\leq H(Y_1)\\
	&\text{over all } p(x).
	\end{split}
	\right.
	\right\}.
	\end{equation}
\end {theorem}

\begin{IEEEproof}[Proof of Theorem~\ref{theorem3}] We present the achievability proof in the following, and the converse proof in Appendix~\ref{AppC}. The region $\mathcal{R}_2$ can be achieved by taking the union of the regions: $\mathcal{R}_{2a}$, $\mathcal{R}_{2b}$, $\mathcal{R}_{2c}$ and $\mathcal{R}_{2d}$ stated below.
\begin{enumerate}[leftmargin=*]
	\item For all $p(x)$ such that $H(Y_1)>H(Y_2)>H(Z)$\newline
	The individual secrecy rate region in this case can be achieved using Theorem~\ref{theorem1} by setting $U=\phi$, $V_1=Y_1$ and $V_2=V=Y_2$. As a result, we have a superposition coding scheme with only the $V^n$ and $V_1^n$ codeword layers. From the proposed coding scheme in Fig.~\ref{fig:2b}, we remove $M_{a}$, $M_{1a}$, $M_{2a}$, $M_{2a1}$, $M_{2a2}$, $M_{2c}$ $D_{2}$, $L_{1}$ and $L_{2}$. Upon removing the corresponding rate terms and constraints, we apply the Fourier-Motzkin procedure to obtain the region $\mathcal{R}_{2a}$ as follows:
	\begin{equation}\label{eq9}
	\left\{
	\begin{split}
	(R_1,R_2)\\ \in \mathbb{R}_+^2
	\end{split}
	\left\vert
	\begin{split}
	&R_1< H(Y_1|Z),\\
	&R_2< H(Y_2|Z),\\
	&R_1+R_2< H(Y_1)\\
	&\text{over all }p(x)\text{ subject to }\\
	&H(Y_1|Z)>0\text{ and } H(Y_1|Y_2)>0.\\
	\end{split}
	\right.
	\right\}.
	\end{equation}
	\item For all $p(x)$ such that $H(Y_1)=H(Y_2)>H(Z)$\newline
	In order to achieve the individual secrecy rate region in this case, from Theorem~\ref{theorem1}, we set $U=\phi$ and $V_2=V_1=V=Y_2=Y_1$. The proposed scheme reduces to the $V^n$ codeword layer. Thus, from the proposed coding scheme in Fig.~\ref{fig:2b}, we are left with $M_{1b}$, $M_{2b}$ and $D$. Upon removing the corresponding rate terms and constraints, we apply the Fourier-Motzkin procedure to obtain the region $\mathcal{R}_{2b}$ as follows:
	\begin{equation}\label{eq10}
	\left\{
	\begin{split}
	(R_1,R_2)\\ \in \mathbb{R}_+^2
	\end{split}
	\left\vert
	\begin{split}
	&R_1< H(Y_1|Z),\\
	&R_2< H(Y_2|Z),\\
	&R_1+R_2< H(Y_1)\\
	&\text{over all }p(x)\text{ subject to }\\
	&H(Y_2|Z)>0.\\
	\end{split}
	\right.
	\right\}.
	\end{equation}
	\item For all $p(x)$ such that $H(Y_1)>H(Z)\geq H(Y_2)$\newline
	For the case in which the eavesdropper is of equal channel strength as the weaker legitimate receiver 2, from Theorem~\ref{theorem1}, we set $V_1=Y_1$ and $V_2=V=U=\phi$. We are essentially disregarding the weaker legitimate receiver 2 and the proposed scheme reduces to the simple wiretap channel coding scheme with only $V_1^n$. Thus, the proposed coding scheme in Fig.~\ref{fig:2b} is reduced to only having $M_{1c}$ and $D_{1}$. Upon removing the corresponding rate terms and constraints, we apply the Fourier-Motzkin procedure to obtain the region $\mathcal{R}_{2c}$ as follows:
	\begin{equation}\label{eq11}
	\left\{
	\begin{split}
	(R_1,R_2)\\ \in \mathbb{R}_+^2
	\end{split}
	\left\vert
	\begin{split}
	&R_1 < H(Y_1|Z),\\
	&R_2 = 0\\
	&\text{over all }p(x).
	\end{split}
	\right.
	\right\}.
	\end{equation}
	\item For all $p(x)$ such that $H(Z)\geq H(Y_1)\geq H(Y_2)$\newline
	When the eavesdropper is not weaker than both legitimate receivers, from Theorem~\ref{theorem1}, we need to set $V_1=V_2=V=U=\phi$. This again indicates that secure communication is not possible and we obtain the region $\mathcal{R}_{1d}=\{(0,0)\}$.
\end{enumerate}
By taking the union of $\mathcal{R}_{2a}$,$\mathcal{R}_{2b}$, $\mathcal{R}_{2c}$ and $\mathcal{R}_{2d}$, we achieve the individual secrecy capacity region of the channel in interest. 
\end {IEEEproof}

\begin{remark}
Revisiting the individual secrecy capacity region for the two-receiver discrete memoryless broadcast channel without receiver side information and with a passive eavesdropper under similar channel degradedness \cite[Theorem 3]{cref1}, we set $Y_1$ to be a deterministic function of $X$. From here, we realize that single-sided receiver side information is not useful in physically degraded deterministic broadcast channel when it is available at receiver 1, which is now the stronger legitimate receiver. The availability of receiver side information does not help in improving the individual secrecy rate region. As a result, when single-sided receiver side information is available at the stronger legitimate receiver, our proposed coding scheme in Fig.~\ref{fig:2b} can also be simplified by removing the $U^n$ codeword, consequently removing $M_{a}$, $M_{1a}$, $M_{2a}$, $M_{2a1}$ and $M_{2a2}$.
\end{remark} 

\section{Conclusion}

In conclusion, we studied the case of the two-receiver discrete memoryless broadcast channel with one-sided receiver side information and with a passive eavesdropper. Under the context of individual secrecy, we presented a coding scheme constructed from a combination of superposition-Marton coding scheme \cite{cref6},\cite{cref5} and secrecy techniques such as the one-time pad \cite{cref8}, Carleial-Hellman secrecy coding \cite{cref3} as well as Wyner serecy coding \cite{cref4}. This scheme is shown to be capacity achieving for some special cases of the physically degraded deterministic broadcast channel. In these cases, we observe that the availability of one-sided receiver side information at the weaker legitimate receiver provides rate region improvements. However, when one-sided receiver side information is available at the stronger legitimate receiver, no gains are obtained since their statistical strength allows full recovery of the intended message. Finally, we note that the implementation of joint secrecy and the derivation of the converse for the broadcast channel with one-sided receiver side information will be covered in future works.

\begin{figure*}[t]
	\normalsize
	\setcounter{MYtempeqncnt}{\value{equation}}
	\setcounter{equation}{17}
	\begin{flalign}
	\mathcal{E}_0&:(U^n(1),V^n(1,1,1,1,d'),V_1^n(1,1,1,1,d',1,d'_1,l_1),V_2^n(1,1,1,1,d',1,1,d'_2,l_2))\notin \mathcal{T}^n_{\epsilon'} \text{ for all } l_2 \text{ and } l_2,\label{cc5}&\\
	\mathcal{E}_{11}&:(U^n(1),V^n(1,1,1,1,d'),V_1^n(1,1,1,1,d',1,d'_1,l'_1),Y_1^n)\notin \mathcal{T}^n_{\epsilon}\label{cc6}&\\
	\mathcal{E}_{12}&:(U^n(1),V^n(1,1,1,1,d'),V_1^n(1,1,1,1,d',m_{1c},d_1,l_1),Y_1^n)\in \mathcal{T}^n_{\epsilon} \text{ for some } m_{1c}\neq 1,d_1 \text{ and } l_1,\label{cc7}&\\
	\mathcal{E}_{13}&:(U^n(1),V^n(1,m_{1b},1,1,d),V_1^n(1,m_{1b},1,1,d,m_{1c},d_1,l_1),Y_1^n)\in \mathcal{T}^n_{\epsilon} \text{ for some } m_{1b}\neq 1,d,m_{1c},d_1 \text{ and } l_1,&\label{cc8}\\
	\mathcal{E}_{14}&:(U^n(m_{a}),V^n(m_{a},m_{1b},1,1,d),V_1^n(m_{a},m_{1b},1,1,d,m_{1c},d_1,l_1),Y_1^n)\in \mathcal{T}^n_{\epsilon}\label{cc9}&\\&\qquad\qquad\qquad\qquad\qquad\qquad\qquad\qquad\qquad\qquad\qquad\qquad\qquad\qquad\qquad\medspace\text{ for some } m_{a}\neq 1,m_{1b},d,m_{1c},d_1 \text{ and } l_1,\nonumber&\\
	\mathcal{E}_{21}&:(U^n(1),V^n(1,1,1,1,d'),V_2^n(1,1,1,1,d',1,1,d'_2,l'_2),Y_2^n)\notin \mathcal{T}^n_{\epsilon}&\label{cc10}\\
	\mathcal{E}_{22}&:(U^n(1),V^n(1,1,1,1,d'),V_2^n(1,1,1,1,d',m_{2a2},m_{2c},d_2,l_2),Y_2^n)\in \mathcal{T}^n_{\epsilon}\label{cc11}\\ &\qquad\qquad\qquad\qquad\qquad\qquad\qquad\qquad\qquad\qquad\qquad\qquad\qquad\qquad\qquad \text{ for some } (m_{2a2},m_{2c})\neq (1,1),d_2 \text{ and } l_2,\nonumber&\\
	\mathcal{E}_{23}&:(U^n(1),V^n(1,m_{1b},m_{2b},m_{2a1},d),V_2^n(1,m_{1b},m_{2b},m_{2a1},d,m_{2a2},m_{2c},d_2,l_2),Y_2^n)\in \mathcal{T}^n_{\epsilon}\label{cc12}\\ &\qquad\qquad\qquad\qquad\qquad\qquad\qquad\qquad\qquad\qquad\qquad\thickspace\text{ for some } (m_{2b},m_{2a1})\neq (1,1),m_{1b},d,m_{2a2},m_{2c},d_2 \text{ and } l_2,\nonumber&\\
	\mathcal{E}_{24}&:(U^n(m_{a}),V^n(m_{a},m_{1b},m_{2b},m_{2a1},d),V_2^n(m_{a},m_{1b},m_{2b},m_{2a1},d,m_{2a2},m_{2c},d_2,l_2),Y_2^n)\in \mathcal{T}^n_{\epsilon}\label{cc13}&\\ &\qquad\qquad\qquad\qquad\qquad\qquad\qquad\qquad\qquad\qquad\qquad\quad\text{ for some } m_{a}\neq 1,m_{1b},m_{2b},m_{2a1},d,m_{2a2},m_{2c},d_2 \text{ and } l_2.\nonumber&
	\end{flalign}
	\setcounter{equation}{\value{MYtempeqncnt}}
	\hrulefill
	\vspace*{4pt}
\end{figure*}

\appendices
\section{}\label{AppA}
In this section, we will present the achievability proof of Theorem~\ref{theorem1}.
\begin{IEEEproof}[Proof of Theorem~\ref{theorem1}]
	\textbf{Rate splitting.} This coding scheme involves splitting $M_i$, for all $i\in\{1,2\}$, into three indepedent message segments, namely $M_{ia}$ at rate $R_{ia}$, $M_{ib}$ at rate $R_{ib}$ and $M_{ic}$ at rate $R_{ic}$, i.e., $M_i=(M_{ia},M_{ib},M_{ic})$ where $M_{ia}\in [1:2^{nR_{ia}}]$, $M_{ib}\in [1:2^{nR_{b}}]$ and $M_{ic}\in [1:2^{nR_{ic}}]$. and $R_i=R_{a}+R_{b}+R_{ic}$.
	This rate splitting imposes the rate relations:
	\begin{flalign}
	R_1=&R_{1a}+R_{1b}+R_{1c},\label{cc1}&\\
	R_2=&R_{2a}+R_{2b}+R_{2c}.\label{cc2}&
	\end{flalign}
	A one-time pad $M_a$ at rate $R_a$ is constructed using the message segments $M_{1a}$ and $M_{2a}$, i.e., $M_a=M_{1a}\oplus M_{2a}$, where $M_{a}\in [1:2^{nR_{a}}]$. This gives the following rate relation:
	\begin{flalign}
	&R_a=R_{1a}=R_{2a}.\label{cc3}&
	\end{flalign}
	We will further split the message segment $M_{2a}$ into $M_{2a1}$ at rate $R_{2a1}$ and $M_{2a2}$ at rate $R_{2a2}$, where $M_{2a1}\in [1:2^{nR_{2a1}}]$, $M_{2a2}\in [1:2^{nR_{2a2}}]$. Both $M_{2a1}$ and $M_{2a2}$ will be inserted into different codeword layers. This gives the following rate relation:
	\begin{flalign}
	&R_{2a}=R_{2a1}+R_{2a2}.\label{cc4}&
	\end{flalign}
	
	\textbf{Codebook generation.} 
	Fix a pmf $p(u)p(v|u)p(v_1,v_2|v)p(x|v_1,v_2)$. Randomly and independently generate $2^{nR_a}$ sequences $u^n(m_a)$, where $m_a=m_{1a}\oplus m_{2a}$, $m_a\in [1:2^{nR_a}]$, each according to $\prod_{i=1}^{n}p_U(u_i)$. 
	
	For each $m_a$, randomly and conditionally independently generate $2^{n[R_{1b}+R_{2b}+R_{2a1}+R_{d}]}$ sequences $v^n(m_a,m_{1b},m_{2b},m_{2a1},d)$, $(m_{1b},m_{2b},m_{2a1},d)\in [1:2^{nR_{1b}}]\times[1:2^{nR_{2b}}]\times[1:2^{nR_{2a1}}]\times[1:2^{nR_{d}}]$, each according to $\prod_{i=1}^{n}p_{V|U}(v_i|u_i(m_a))$. 
	
	For each $(m_a,m_{1b},m_{2b},m_{2a1},d)$, randomly and conditionally independently generate $2^{n[R_{1c}+R_{d1}+R_{l1}]}$ sequences $v^n_1(m_a,m_{1b},m_{2b},m_{2a1},d,m_{1c},d_1,l_1)$, $(m_{1c},d_1,l_1)\in [1:2^{nR_{1c}}]\times[1:2^{nR_{d1}}]\times[1:2^{nR_{l1}}]$, each according to $\prod_{i=1}^{n}p_{V_1|V}(v_{1i}|v_i(m_a,m_{1b},m_{2b},m_{2a1},d))$. 
	
	Similarly, for each $(m_a,m_{1b},m_{2b},m_{2a1},d)$, randomly and conditionally independently generate $2^{n[R_{2a2}+R_{2c}+R_{d2}+R_{l2}]}$ sequences $v^n_2(m_a,m_{1b},m_{2b},m_{2a1},d,m_{2a2},m_{2c},d_2,l_2)$, $(m_{2a2},m_{2c},d_2,l_2)\in [1:2^{nR_{2a2}}]\times[1:2^{nR_{2c}}]\times[1:2^{nR_{d2}}]\times[1:2^{nR_{l2}}]$, each according to $\prod_{i=1}^{n}p_{V_2|V}(v_{2i}|v_i(m_a,m_{1b},m_{2b},m_{2a1},d))$. This codebook is revealed to all parties (including the eavesdropper).
	
	\textbf{Encoding.} To send $(m_1,m_2)$, the encoder chooses $u^n(m_a)$, where $m_a\triangleq m_{1a}\oplus m_{2a}$. Given $u^n(m_a)$, the encoder independently generates $d$ with equal probability over $[1:2^{nR_{d}}]$ and find $v^n(m_a,m_{1b},m_{2b},m_{2a1},d)$. 
	
	Given $v^n(m_a,m_{1b},m_{2b},m_{2a1},d)$, the encoder independently generates $d_1$ and $d_2$ with equal probability over $ [1:2^{nR_{d1}}]$ and $[1:2^{nR_{d2}}]$ respectively before finding an index pair $(l_1,l_2)$ such that $(v^n_1(\cdot,l_1),v^n_2(\cdot,l_2))\in \mathcal{T}^n_{\epsilon'}$. If there is more than one such pair, the encoder chooses an arbitrary one among those. If no such pair exists, choose $(l_1,l_2)=(1,1)$. 
	
	Lastly, given the chosen jointly typical pair $(v^n_1,v^n_2)$, it then randomly generates the codeword $X^n(m_1,m_2)\sim\prod_{i=1}^{n}p_{X|V_1,V_2}(x_i|v_{1i}(\cdot),v_{2i}(\cdot))$ and transmits it. 
	
	\textbf{Decoding.}
	Let $\epsilon>\epsilon'$. Receiver 1 declares that $(\hat{m}_a,\hat{m}_{1b},\hat{m}_{1c})$ is sent if it is the unique tuple such that $(u^n(\cdot),v^n(\cdot,d),v^n_1(\cdot,d,\cdot,d_1,l_1),y^n_1)\in\mathcal{T}^n_{\epsilon}$ for some $(d,d_1,l_1)$; otherwise it declares an error. Since receiver 1 knows $M_2$ as receiver message side informaiton, it decodes $(u^n,v^n,v_1^n)$ over a set of $2^{n(R_a+R_{1b}+R_d+R_{1c}+R_{d1}+R_{l1})}$ candidates. 
	
	On the other hand, receiver 2 declares that $(\hat{m}_a,\hat{m}_{2b},\hat{m}_{2a1},\hat{m}_{2a2},\hat{m}_{2c})$ is sent if it is the unique tuple such that $(u^n(\cdot),v^n(\cdot,m_{1b},d),v^n_2(\cdot,m_{1b},d,\cdot,d_2,l_2),y^n_2)\in\mathcal{T}^n_{\epsilon}$ for some $(m_{1b},d,d_2,l_2)$; otherwise it declares an error. In contrast to receiver 1, receiver 2 does not have receiver message side informaiton and it decodes $(u^n,v^n,v_2^n)$ over all $2^{n(R_a+R_{1b}+R_{2b}+R_d+R_{2a}+R_{2c}+R_{d2}+R_{l2})}$ candidates.
	
	\textbf{Analysis of the probability of error.} 
	Assume without loss of generality that the transmitted messages are equal to one and $(d,d_1,d_2,l_1,l_2)=(d',d'_1,d'_2,l'_1,l'_2)$. Receiver 1 makes an error only if one or more of the error events (\ref{cc5})--(\ref{cc9}) occur. On the other hand, receiver 2 makes an error only if one or more of the error events (\ref{cc5}), (\ref{cc10})--(\ref{cc13}) occur.
	
	\addtocounter{equation}{9}
	Considering the error events, we apply the mutual covering lemma and the packing lemma to obtain the achievability conditions:
	\begin{flalign}
	R_{l1}+R_{l2}&>I(V_1;V_2|U,V),\label{cc14}&\\
	R_1+R_d+R_{d1}+R_{l1}&<I(U,V,V_1;Y_1),\label{cc15}&\\
	R_1-R_a+R_d+R_{d1}+R_{l1}&<I(V,V_1;Y_1|U),\label{cc16}&\\
	R_{1c}+R_{d1}+R_{l1}&<I(V_1;Y_1|U,V),\label{cc17}&\\
	R_2+R_{1b}+R_d+R_{2a}&\nonumber\\+R_{d2}+R_{l2}&<I(U,V,V_2;Y_2),\label{cc18}&\\
	R_2-R_a+R_{1b}+R_d+R_{2a}&\nonumber\\+R_{d2}+R_{l2}&<I(V,V_2;Y_2|U),\label{cc19}&\\
	R_{2a2}+R_{2c}+R_{d2}+R_{l2}&<I(V_2;Y_2|U,V).\label{cc20}&
	\end{flalign}
	
	\textbf{Analysis of individual secrecy.} In order to ensure the individual secrecy of both messages, we need to satisfy $R_{\text{L},i}^{(n)}\leq \tau_n\text{, for all }i\in\{1,2\}$ as stated in (\ref{sec}). Here, we show that $I(M_1;Z^n)\leq n\tau_n$ and $I(M_2;Z^n)\leq n\tau_n$, or alternatively $I(M_1;Z^n)+I(M_2;Z^n)\leq 2n\tau_n$.
	
	For the individual secrecy of $M_1$, we have
	\begin{flalign}
	&I(M_1;Z^n)&\nonumber\\
	&=I(M_{1a},M_{1b},M_{1c};Z^n)\nonumber&\\
	&=I(M_{1a};Z^n)+I(M_{1b},M_{1c};Z^n|M_{1a})\nonumber&\\
	&=I(M_{1a};Z^n)+H(M_{1b},M_{1c}|M_{1a})-H(M_{1b},M_{1c}|Z^n,M_{1a})\nonumber&\\
	&=n[R_{1b}+R_{1c}]+I(M_{1a};Z^n)-H(M_{1b},M_{1c}|Z^n,M_{1a})\nonumber&\\
	&\leq n[R_{1b}+R_{1c}]+I(M_{1a};M_a,Z^n)-H(M_{1b},M_{1c}|Z^n,M_{1a})\nonumber&\\
	&= n[R_{1b}+R_{1c}]+I(M_{1a};Z^n|M_a)-H(M_{1b},M_{1c}|Z^n,M_{1a})\nonumber&\\
	&\leq n[R_{1b}+R_{1c}]+I(M_{1a},M_{2a};Z^n|M_a)-H(M_{1b},M_{1c}|Z^n,M_{1a})\nonumber&\\
	&= n[R_{1b}+R_{1c}]+I(M_{2a};Z^n|M_a)-H(M_{1b},M_{1c}|Z^n,M_{1a})\nonumber&\\
	&= n[R_{1b}+R_{1c}]+H(M_{2a}|M_a)-H(M_{2a}|Z^n,M_a)\nonumber&\\&\quad-H(M_{1b},M_{1c}|Z^n,M_{1a})\nonumber&\\
	&= n[R_{1b}+R_{1c}+R_{2a}]-H(M_{2a}|Z^n,M_a)\nonumber&\\&\quad-H(M_{1b},M_{1c}|Z^n,M_{1a})\nonumber&\\
	&\leq  n[R_{1b}+R_{1c}+R_{2a}]-H(M_{2a}|Z^n,M_a)\nonumber&\\&\quad-H(M_{1b},M_{1c}|Z^n,M_a,M_{1a})\nonumber&\\	
	&= n[R_{1b}+R_{1c}+R_{2a}]-H(M_{2a}|Z^n,M_a)\nonumber&\\&\quad-H(M_{1b},M_{1c}|Z^n,M_a)\nonumber&\\
	&= n[R_{1b}+R_{1c}+R_{2a}]-H(M_{2a1},M_{2a2}|Z^n,M_a)\nonumber&\\&\quad-H(M_{1b},M_{1c}|Z^n,M_a)\nonumber&\\
	&=n[R_{1b}+R_{1c}+R_{2a}] \nonumber&\\&\quad - H(M_{1b},M_{2b},M_{2a1},D,M_{2a2},M_{2c},D_2|Z^n,M_a)\nonumber&\\&\quad + H(M_{1b},M_{2b},D,M_{2c},D_2|Z^n,M_a,M_{2a1},M_{2a2})\nonumber&\\&\quad- H(M_{1b},M_{2b},D,M_{1c},D_1|Z^n,M_a)\nonumber& \\&\quad + H(M_{2b},D,D_1|Z^n,M_a,M_{1b},M_{1c})\nonumber&\\
	&=n[R_{1b}+R_{1c}+R_{2a}] \nonumber&\\&\quad - H(M_{1b},M_{2b},M_{2a1},D|Z^n,M_a) \nonumber&\\&\quad- H(M_{2a2},M_{2c},D_2|Z^n,M_a,M_{1b},M_{2b},M_{2a1},D)\nonumber&\\&\quad + H(M_{1b},M_{2b},D|Z^n,M_a,M_{2a1},M_{2a2}) \nonumber&\\&\quad+ H(M_{2c},D_2|Z^n,M_a,M_{2a1},M_{2a2},M_{1b},M_{2b},D)\nonumber&\\&\quad
	- H(M_{1b},M_{2b},D|Z^n,M_a)\nonumber& \\&\quad - H(M_{1c},D_1|Z^n,M_a,M_{1b},M_{2b},D)\nonumber& \\&\quad + H(M_{2b},D|Z^n,M_a,M_{1b},M_{1c})\nonumber& \\&\quad + H(D_1|Z^n,M_a,M_{1b},M_{1c},M_{2b},D)\nonumber&\\
	&\leq n[R_{1b}+R_{1c}+R_{2a}] \nonumber&\\&\quad - H(M_{1b},M_{2b},M_{2a1},D|Z^n,M_a) \nonumber&\\&\quad- H(M_{2a2},M_{2c},D_2|Z^n,M_a,M_{1b},M_{2b},M_{2a1},D)\nonumber&\\&\quad + H(M_{1b},M_{2b},D|Z^n,M_a,M_{2a1}) \nonumber&\\&\quad+ H(M_{2c},D_2|Z^n,M_a,M_{1b},M_{2b},M_{2a1},D,M_{2a2})\nonumber&\\&\quad
	- H(M_{1b},M_{2b},D|Z^n,M_a)\nonumber& \\&\quad - H(M_{1c},D_1|Z^n,M_a,M_{1b},M_{2b},D)\nonumber& \\&\quad + H(M_{2b},D|Z^n,M_a,M_{1b})\nonumber& \\&\quad + H(D_1|Z^n,M_a,M_{1b},M_{2b},D,M_{1c})\label{cc21}
	\end{flalign}
	
	We establish a lower bound on the equivocation term $H(M_{1b},M_{2b},D|Z^n,M_a)$ in (\ref{cc21}).
	\begin{flalign}
	&H(M_{1b},M_{2b},D|Z^n,M_a)\nonumber&\\
	&=H(M_{1b},M_{2b},D|M_a) - I(M_{1b},M_{2b},D;Z^n|M_a)\nonumber&\\
	&=H(M_{1b},M_{2b},D) - I(M_{1b},M_{2b},D;Z^n|M_a)\nonumber&\\
	&=n[R_{1b}+R_{2b}+R_d] - I(M_{1b},M_{2b},D;Z^n|M_a)\nonumber&\\
	&=n[R_{1b}+R_{2b}+R_d] - I(V^n,M_{1b},M_{2b},D;Z^n|U^n,M_a)\nonumber&\\
	&\overset{\text{(a)}}{=}n[R_{1b}+R_{2b}+R_d] - I(V^n;Z^n|U^n)\nonumber&\\
	&\overset{\text{(b)}}{\geq}n[R_{1b}+R_{2b}+R_d] - nI(V;Z|U) - n\delta_1(\tau)\label{cc22}&
	\end{flalign}
	where (a) follows since $M_{a}$$\quad \rightarrow U^n\rightarrow Z^n$ and $(M_{1b},M_{2b},D)\rightarrow V^n\rightarrow Z^n$ form Markov chains and (b) follows since $I(V^n;Z^n|U^n)\leq nI(V;Z|U)+n\delta_1(\tau)$ as in \cite[Lemma 3]{Liu_Maric_Spasojevic_Yates08}.
	
	We establish a lower bound on the equivocation term $H(M_{1c},D_1|Z^n,M_a,M_{1b},M_{2b},D)$ in (\ref{cc21}).
	\begin{flalign}
	&H(M_{1c},D_1|Z^n,M_a,M_{1b},M_{2b},D)&\nonumber\\
	&=H(M_{1c},D_1|M_a,M_{1b},M_{2b},D)\nonumber&\\& \quad - I(M_{1c},D_1;Z^n|M_a,M_{1b},M_{2b},D)\nonumber&\\
	&=H(M_{1c},D_1) - I(M_{1c},D_1;Z^n|M_a,M_{1b},M_{2b},D)\nonumber&\\
	&=n[R_{1c}+R_{d1}] - I(M_{1c},D_1;Z^n|M_a,M_{1b},M_{2b},D)\nonumber&\\
	&=n[R_{1c}+R_{d1}]\nonumber&\\& \quad - I(V_1^n,M_{1c},D_1;Z^n|U^n,M_a,V^n,M_{1b},M_{2b},D)\nonumber&\\
	&\overset{\text{(c)}}{=}n[R_{1c}+R_{d1}] - I(V_1^n;Z^n|U^n,V^n)\nonumber&\\
	&\geq n[R_{1c}+R_{d1}] - nI(V_1;Z|U,V) - n\delta_2(\tau)\label{cc23}&
	\end{flalign}
	where (c) follows since $M_a \rightarrow U^n \rightarrow Z^n$, $(M_{1b},M_{2b},D) \rightarrow V^n \rightarrow Z^n$ and $(M_{1c},D_1)\rightarrow (V^n,V_1^n)\rightarrow Z^n$ form Markov chains. The third Markov chain can be proven using the functional	dependence graph \cite[Definition A.1]{Kramer08} and the fact that $V^n$ can be retrieved by knowing $V_1^n$. 
	
	We establish a lower bound on the equivocation term $H(M_{1b},M_{2b},M_{2a1},D|Z^n,M_a)$ in (\ref{cc21}).
	\begin{flalign}
	&H(M_{1b},M_{2b},M_{2a1},D|Z^n,M_{a})\nonumber&\\
	&=H(M_{1b},M_{2b},M_{2a1},D|M_a) - I(M_{1b},M_{2b},M_{2a1},D;Z^n|M_a)\nonumber&\\
	&=H(M_{1b},M_{2b},M_{2a1},D) - I(M_{1b},M_{2b},M_{2a1},D;Z^n|M_a)\nonumber&\\
	&=n[R_{1b}+R_{2b}+R_{2a1}+R_d] - I(M_{1b},M_{2b},M_{2a1},D;Z^n|M_a)\nonumber&\\
	&=n[R_{1b}+R_{2b}+R_{2a1}+R_d]\nonumber&\\&\quad - I(V^n,M_{1b},M_{2b},M_{2a1},D;Z^n|U^n,M_a)\nonumber&\\
	&\overset{\text{(d)}}{=}n[R_{1b}+R_{2b}+R_{2a1}+R_d] - I(V^n;Z^n|U^n)\nonumber&\\
	&\geq n[R_{1b}+R_{2b}+R_{2a1}+R_d] - nI(V;Z|U) - n\delta_3(\tau)\label{cc24}&
	\end{flalign}
	where (d) follows since $M_a$$\quad \rightarrow U^n\rightarrow Z^n$ and $(M_{1b},M_{2b},M_{2a1},D)\rightarrow V^n\rightarrow Z^n$ form Markov chains.
	
	We establish a lower bound on the equivocation term $H(M_{2a2},M_{2c},D_2|Z^n,M_a,M_{1b},M_{2b},M_{2a1},D)$ in (\ref{cc21}).
	\begin{flalign}
	&H(M_{2a2},M_{2c},D_2|Z^n,M_{a},M_{1b},M_{2b},M_{2a1},D)&\nonumber\\
	&=H(M_{2a2},M_{2c},D_2|M_{a},M_{1b},M_{2b},M_{2a1},D)\nonumber&\\& \quad - I(M_{2a2},M_{2c},D_2;Z^n|M_{a},M_{1b},M_{2b},M_{2a1},D)\nonumber&\\
	&=H(M_{2a2},M_{2c},D_2)\nonumber&\\& \quad - I(M_{2a2},M_{2c},D_2;Z^n|M_{a},M_{1b},M_{2b},M_{2a1},D)\nonumber&\\
	&=n[R_{2a2}+R_{2c}+R_{d2}]\nonumber&\\& \quad - I(M_{2a2},M_{2c},D_2;Z^n|M_{a},M_{1b},M_{2b},M_{2a1},D)\nonumber&\\
	&=n[R_{2a2}+R_{2c}+R_{d2}]\nonumber - I(V_2^n,M_{2a2},M_{2c},D_2;Z^n|U^n,&\\& \quad M_{a},V^n,M_{1b},M_{2b},M_{2a1},D)\nonumber&\\
	&\overset{\text{(e)}}{=}n[R_{2a2}+R_{2c}+R_{d2}] - I(V_2^n;Z^n|U^n,V^n)\nonumber&\\
	&\geq n[R_{2a2}+R_{2c}+R_{d2}] - nI(V_2;Z|U,V) - n\delta_4(\tau)\label{cc25}&
	\end{flalign}
	where (e) follows since $M_a$$\quad \rightarrow U^n \rightarrow Z^n$, $(M_{1b},M_{2b},M_{2a1},D) \rightarrow V^n \rightarrow Z^n$ and $(M_{2a2},M_{2c},D_2)\rightarrow (V^n,V_2^n)\rightarrow Z^n$ form Markov chains. The third Markov chain can be proven using the functional dependence graph \cite[Definition A.1]{Kramer08} and the fact that $V^n$ can be retrieved by knowing $V_2^n$.
	
	We establish an upper bound on the equivocation term $H(M_{2b},D|Z^n,M_a,M_{1b})$ in (\ref{cc21}). By \cite[Lemma 22.1]{cref6}, 
	\vspace{-0.75pt}
	\begin{flalign}
	&H(M_{2b},D|Z^n,M_a,M_{1b})&\nonumber\\&\leq n[R_{2b}+R_d-I(V;Z|U)]+n\delta_5(\tau)\label{cc26}&
	\end{flalign}
	if 
	\begin{flalign}
	R_{2b}+R_d \geq I(V;Z|U)\label{cc27}
	\end{flalign}
	
	We establish an upper bound on the equivocation term $H(D_1|Z^n,M_a,M_{1b},M_{2b},D,M_{1c})$ in (\ref{cc21}). By \cite[Lemma 22.1]{cref6},
	\vspace{-0.75pt}
	\begin{flalign}
	&H(D_1|Z^n,M_a,M_{1b},M_{2b},D,M_{1c})&\nonumber\\&\leq n[R_{d1}-I(V_1;Z|U,V)]+n\delta_6(\tau)\label{cc28}&
	\end{flalign}
	if 
	\begin{flalign}
	R_{d1} \geq I(V_1;Z|U,V)\label{cc29}
	\end{flalign}
	
	We establish an upper bound on the equivocation term $H(M_{1b},M_{2b},D|Z^n,M_a,M_{2a1})$ in (\ref{cc21}). By \cite[Lemma 22.1]{cref6},
	\vspace{-0.75pt}
	\begin{flalign}
	&H(M_{1b},M_{2b},D|Z^n,M_a,M_{2a1})&\nonumber\\&\leq n[R_{1b}+R_{2b}+R_d-I(V;Z|U)]+n\delta_7(\tau)\label{cc30}&
	\end{flalign}
	if 
	\begin{flalign}
	R_{1b}+R_{2b}+R_d \geq I(V;Z|U)\label{cc31}
	\end{flalign}
		
	We establish an upper bound on the equivocation term $H(M_{2c},D_2|Z^n,M_a,M_{1b},M_{2b},M_{2a1},D,M_{2a2})$ in (\ref{cc21}). By \cite[Lemma 22.1]{cref6},
	\vspace{-0.75pt}
	\begin{flalign}
	&H(M_{2c},D_2|Z^n,M_a,M_{1b},M_{2b},M_{2a1},D,M_{2a2})&\nonumber\\&\leq n[R_{2c}+R_{d2}-I(V_2;Z|U,V)]+n\delta_8(\tau)\label{cc32}&
	\end{flalign}
	if 
	\begin{flalign}
	R_{2c}+R_{d2} \geq I(V_2;Z|U,V)\label{cc33}
	\end{flalign}
	
	For the individual secrecy of $M_2$, we have
	\begin{flalign}
	&I(M_2;Z^n)\nonumber&\\
	&=I(M_{2a1},M_{2a2},M_{2b},M_{2c};Z^n)\nonumber&\\
	&=H(M_{2a1},M_{2a2},M_{2b},M_{2c}) \nonumber&\\&\quad- H(M_{2a1},M_{2a2},M_{2b},M_{2c}|Z^n)\nonumber&\\
	&=n[R_{2a1}+R_{2a2}+R_{2b}+R_{2c}] \nonumber&\\&\quad- H(M_{2a1},M_{2a2},M_{2b},M_{2c}|Z^n)\nonumber&\\
	&=n[R_{2a}+R_{2b}+R_{2c}]\nonumber&\\&\quad - H(M_{a},M_{1b},M_{2b},M_{2a1},M_{2a2},D,M_{2c},D_2|Z^n)\nonumber&\\&\quad + H(M_{a},M_{1b},D,D_2|Z^n,M_{2a1},M_{2a2},M_{2b},M_{2c})\nonumber&\\
	&=n[R_{2a}+R_{2b}+R_{2c}]\nonumber&\\&\quad - H(M_{a}|Z^n) - H(M_{1b},M_{2b},M_{2a1},D|Z^n,M_{a})\nonumber&\\&\quad - H(M_{2a2},M_{2c},D_2|Z^n,M_{a},M_{1b},M_{2b},M_{2a1},D)\nonumber&\\&\quad + H(M_{a}|Z^n,M_{2a1},M_{2a2},M_{2b},M_{2c})\nonumber&\\&\quad + H(M_{1b},D|Z^n,M_{a},M_{2a1},M_{2a2},M_{2b},M_{2c})\nonumber&\\&\quad +
	H(D_2|Z^n,M_{a},M_{2a1},M_{2a2},M_{1b},M_{2b},D,M_{2c})\nonumber&\\
	&\leq n[R_{2a}+R_{2b}+R_{2c}] - H(M_{a}|Z^n) + H(M_{a}|Z^n)\nonumber&\\&\quad - H(M_{1b},M_{2b},M_{2a1},D|Z^n,M_{a})\nonumber&\\&\quad- H(M_{2a2},M_{2c},D_2|Z^n,M_{a},M_{1b},M_{2b},M_{2a1},D)\nonumber&\\&\quad + H(M_{1b},D|Z^n,M_{a},M_{2a1},M_{2b})\nonumber&\\&\quad +
	H(D_2|Z^n,M_{a},M_{2a1},M_{2a2},M_{1b},M_{2b},D,M_{2c})\nonumber&\\
	&=n[R_{2a}+R_{2b}+R_{2c}] - H(M_{1b},M_{2b},M_{2a1},D|Z^n,M_{a})\nonumber&\\&\quad - H(M_{2a2},M_{2c},D_2|Z^n,M_{a},M_{1b},M_{2b},M_{2a1},D)\nonumber&\\&\quad + H(M_{1b},D|Z^n,M_{a},M_{2a1},M_{2b})\nonumber&\\&\quad +
	H(D_2|Z^n,M_{a},M_{2a1},M_{2a2},M_{1b},M_{2b},D,M_{2c})\label{cc34}&
	\end{flalign}
	
	We establish a lower bound on the equivocation term $H(M_{1b},M_{2b},M_{2a1},D|Z^n,M_{a})$ in (\ref{cc34}) and obtain the same bound as in (\ref{cc24}). Replacing $\delta_3(\tau)$ with $\delta_9(\tau)$, we have 
	\begin{flalign}
	&H(M_{1b},M_{2b},M_{2a1},D|Z^n,M_{a})\nonumber&\\
	&\geq n[R_{1b}+R_{2b}+R_{2a1}+R_d] - nI(V;Z|U) - n\delta_9(\tau)\label{cc35}&
	\end{flalign}
	
	We establish a lower bound on the equivocation term $H(M_{2a2},M_{2c},D_2|Z^n,M_{a},M_{1b},M_{2b},M_{2a1},D)$ in (\ref{cc34}).
	\begin{flalign}
	&H(M_{2a2},M_{2c},D_2|Z^n,M_{a},M_{1b},M_{2b},M_{2a1},D)\nonumber&\\
	&=H(M_{2a2},M_{2c},D_2|M_{a},M_{1b},M_{2b},M_{2a1},D)\nonumber&\\& \quad - I(M_{2a2},M_{2c},D_2;Z^n|M_{a},M_{1b},M_{2b},M_{2a1},D)\nonumber&\\
	&=H(M_{2a2},M_{2c},D_2)\nonumber&\\& \quad - I(M_{2a2},M_{2c},D_2;Z^n|M_{a},M_{1b},M_{2b},M_{2a1},D)\nonumber&\\
	&=n[R_{2a2}+R_{2c}+R_{d2}]\nonumber&\\& \quad - I(M_{2a2},M_{2c},D_2;Z^n|M_{a},M_{1b},M_{2b},M_{2a1},D)\nonumber&\\
	&=n[R_{2a2}+R_{2c}+R_{d2}]\nonumber - I(V_2^n,M_{2a2},M_{2c},D_2;Z^n|U^n,&\\& \quad M_{a},V^n,M_{1b},M_{2b},M_{2a1},D)\nonumber&\\
	&\overset{\text{(f)}}{=}n[R_{2a2}+R_{2c}+R_{d2}] - I(V_2^n;Z^n|U^n,V^n)\nonumber&\\
	&\geq n[R_{2a2}+R_{2c}+R_{d2}] - nI(V_2;Z|U,V) - n\delta_{10}(\tau)\label{cc36}&
	\end{flalign}
	where (f) follows since $M_a$$\quad \rightarrow U^n \rightarrow Z^n$, $(M_{1b},M_{2b},M_{2a1},D) \rightarrow V^n \rightarrow Z^n$ and $(M_{2a2},M_{2c},D_2)\rightarrow (V^n,V_2^n)\rightarrow Z^n$ form Markov chains. The third Markov chain can be proven using the functional dependence graph \cite[Definition A.1]{Kramer08} and the fact that $V^n$ can be retrieved by knowing $V_2^n$.
	
	We establish an upper bound on the equivocation term $H(M_{1b},D|Z^n,M_{a},M_{2a1},M_{2b})$ in (\ref{cc34}). By \cite[Lemma 22.1]{cref6}, 
	\begin{flalign}
	&H(M_{1b},D|Z^n,M_{a},M_{2a1},M_{2b})&\nonumber\\&\leq n[R_{1b}+R_d-I(V;Z|U)]+n\delta_{11}(\tau)\label{cc37}&
	\end{flalign}
	if 
	\begin{flalign}
	R_{1b}+R_d \geq I(V;Z|U)\label{cc38}
	\end{flalign}
	
	We establish an upper bound on the equivocation term $H(D_2|Z^n,M_{a},M_{2a1},M_{2a2},M_{1b},M_{2b},D,M_{2c})$ in (\ref{cc34}). By \cite[Lemma 22.1]{cref6},
	\begin{flalign}
	&H(D_2|Z^n,M_{a},M_{2a1},M_{2a2},M_{1b},M_{2b},D,M_{2c})&\nonumber\\&\leq n[R_{d2}-I(V_2;Z|U,V)]+n\delta_{12}(\tau)\label{cc39}&
	\end{flalign}
	if 
	\begin{flalign}
	R_{d2} \geq I(V_2;Z|U,V)\label{cc40}
	\end{flalign}
	
	Upon substituting (\ref{cc22})--(\ref{cc25}), (\ref{cc26}), (\ref{cc28}), (\ref{cc30}), (\ref{cc32}) into (\ref{cc21}) and substituting (\ref{cc35})--(\ref{cc37}), (\ref{cc39}) into (\ref{cc34}), we obtain the sum of (\ref{cc21}) and (\ref{cc34}) as
	\begin{flalign*}
	&I(M_1;Z^n)+I(M_2;Z^n)&\\
	&\leq n[R_{1b}+R_{1c}+R_{2a}] - n[R_{1b}+R_{2b}+R_d] + nI(V;Z|U)& \\&\quad  + n\delta_1(\tau)- n[R_{1c}+R_{d1}] + nI(V_1;Z|U,V) + n\delta_2(\tau)& \\&\quad -n[R_{1b}+R_{2b}+R_{2a1}+R_d] + nI(V;Z|U) + n\delta_3(\tau)& \\&\quad -n[R_{2a2}+R_{2c}+R_{d2}] + nI(V_2;Z|U,V) + n\delta_4(\tau)&  \\&\quad + n[R_{2b}+R_d-I(V;Z|U)]+n\delta_5(\tau)& \\&\quad + n[R_{d1}-I(V_1;Z|U,V)]+n\delta_6(\tau)& \\&\quad + n[R_{1b}+R_{2b}+R_d-I(V;Z|U)]+n\delta_7(\tau)& \\&\quad +
	n[R_{2c}+R_{d2}-I(V_2;Z|U,V)]+n\delta_8(\tau)& \\
	&\quad
	+n[R_{2a}+R_{2b}+R_{2c}] - n[R_{1b}+R_{2b}+R_{2a1}+R_d]& \\
	&\quad + nI(V;Z|U) + n\delta_{9}(\tau) - n[R_{2a2}+R_{2c}+R_{d2}]&\\&\quad + nI(V_2;Z|U,V) + n\delta_{10}(\tau) + n[R_{1b}+R_d-I(V;Z|U)]&\\&\quad+n\delta_{11}(\tau) +
	n[R_{d2}-I(V_2;Z|U,V)]+n\delta_{12}(\tau)&\\
	&= n\sum_{i=1}^{12}\delta_i(\tau)&\\
	&\overset{\text{(g)}}{=}2n\tau_n&
	\end{flalign*}
	where (g) follows by taking $2\tau_n=\sum_{i=1}^{12}\delta_i(\tau)$.
	
	\textbf{Achievable rate region:} In short, the individual secrecy rate region can be obtained from the following constraints on: 
	\begin{itemize}
		\item the non-negativity of rates; 
		\item the rate relations imposed by rate splitting (\ref{cc1})--(\ref{cc4});
		\item the achievability conditions (\ref{cc14})--(\ref{cc20}); and
		\item the individual secrecy conditions (\ref{cc27}), (\ref{cc29}), (\ref{cc38}), (\ref{cc40}). Note that the conditions (\ref{cc31}) and (\ref{cc33}) are not included since both are redundant.
	\end{itemize}
	Applying the Fourier-Motzkin procedure \cite[Appendix D]{cref6} to eliminate the terms $R_a$, $R_{1b}$, $R_{1c}$, $R_{2a}$, $R_{2b}$, $R_{2c}$ , $R_{d}$, $R_{d1}$, $R_{d2}$, $R_{l1}$ and $R_{l2}$ we obtain the individual secrecy rate region $\mathcal{R}$ in Theorem~\ref{theorem1}. 
\end{IEEEproof}

\section{}\label{AppB}
In this section, we will present the converse proof of Theorem~\ref{theorem2}.
\begin{IEEEproof}[Proof of Theorem~\ref{theorem2}] \textit{(Converse)}
By Fano's inequality \cite[p. 19]{cref6}, we have the reliability constraints
\begin{flalign}
H(M_1|Y_1^n,M_2)\leq n\epsilon_n\label{aa1},\\
H(M_2|Y_2^n)\leq n\epsilon_n\label{aa2},
\end{flalign}
where $\epsilon_n\rightarrow 0$ as $n \rightarrow \infty$.
From (\ref{aa1}) and the fact that $Y_1$ is a degraded version of $Y_2$, we have
\begin{flalign} 
H(M_1|Y_2^n,M_2)\leq H(M_1|Y_1^n,M_2)\leq n\epsilon_n
\end{flalign}
and consequently
\begin{flalign} 
H(M_1|Y_2^n,M_2)\leq n\epsilon_n.\label{aa3}
\end{flalign}
With this, we establish 
\begin{flalign}
H(M_1|Y_2^n)&\leq H(M_1|Y_2^n)+H(M_2|Y_2^n,M_1)\nonumber\\
&=H(M_1,M_2|Y_2^n)\nonumber\\
&=H(M_2|Y_2^n)+H(M_1|Y_2^n,M_2)\nonumber\\
&\overset{\text{(a)}}{\leq}n\epsilon_n+H(M_1|Y_2^n,M_2)\nonumber\\
&\overset{\text{(b)}}{\leq}2n\epsilon_n,\label{aa4}
\end{flalign}
where (a) follows from (\ref{aa2}) and (b) follows from (\ref{aa3}). At the same time, due to individual secrecy, we have the secrecy constraints 
\begin{flalign}
I(M_1;Z^n)\leq n\tau_n\label{aa5},\\
I(M_2;Z^n)\leq n\tau_n\label{aa6}.
\end{flalign}

Using (\ref{aa1}), (\ref{aa2}) and (\ref{aa3})--(\ref{aa6}), we establish the following.
\begin{flalign*}
nR_1&=H(M_1)&\\
&=I(M_1;Y_1^n|M_2)+H(M_1|Y_1^n,M_2)&\\
&\overset{\text{(c)}}{\leq}I(M_1;Y_1^n|M_2)+n\epsilon_n&\\
&=\sum_{i=1}^{n}I(M_1;Y_{1i}|Y_1^{i-1},M_2)+n\epsilon_n&\\
&\leq \sum_{i=1}^{n}I(M_1,M_2,Y_1^{i-1};Y_{1i})+n\epsilon_n&\\
&\overset{\text{(d)}}{\leq} \sum_{i=1}^{n}I(X_i;Y_{1i})+n\epsilon_n&\\
&\overset{\text{(e)}}{\leq} nI(X;Y_{1})+n\epsilon_n&\\
&\overset{\text{(f)}}{=} nH(Y_1)+n\epsilon_n&
\end{flalign*}
where (c) follows from (\ref{aa1}), (d) follows from the Markov chain $(M_1,M_2,Y_1^{i-1}) \rightarrow X_i \rightarrow Y_{1i}$, (e) follows from the time-sharing argument and (f) follows from the fact that $H(Y_1|X)=0$ since $Y_1$ is a deterministic function of $X$.

Besides, we have
\begin{flalign*}
nR_1&=H(M_1)&\\
&\leq I(M_1;Y_1^n|M_2)+n\epsilon_n&\\
&= I(M_1;Y_1^n|M_2)-I(M_1;Z^n|M_2)+I(M_1;Z^n|M_2)&\\&\quad +n\epsilon_n&\\
&\leq I(M_1;Y_1^n,Z^n|M_2)-I(M_1;Z^n|M_2)&\\&\quad+I(M_1;Z^n|M_2)+n\epsilon_n&\\
&= I(M_1;Y_1^n|Z^n,M_2)+I(M_1;Z^n|M_2)+n\epsilon_n&\\
&=\sum_{i=1}^{n}I(M_1;Y_{1i}|Y_{1}^{i+1},Z^n,M_2)+I(M_1;Z^n|M_2)+n\epsilon_n&\\
&\leq \sum_{i=1}^{n}I(M_1,M_2,Y_{1}^{i+1},Z^{i-1},Z_{i+1}^{n};Y_{1i}|Z_i)&\\&\quad+I(M_1;Z^n|M_2)+n\epsilon_n&\\
&\overset{\text{(g)}}{\leq} \sum_{i=1}^{n}I(X_i;Y_{1i}|Z_i)+I(M_1;Z^n|M_2)+n\epsilon_n&\\
&\overset{\text{(h)}}{\leq} nI(X;Y_{1}|Z)+I(M_1;Z^n|M_2)+n\epsilon_n&\\
&\overset{\text{(i)}}{=} nH(Y_1|Z)+I(M_1;Z^n|M_2)+n\epsilon_n&\\
&\leq nH(Y_1|Z)+I(M_1,M_2;Z^n)+n\epsilon_n&\\
&= nH(Y_1|Z)+I(M_1;Z^n)+I(M_2;Z^n|M_1)+n\epsilon_n&\\
&\overset{\text{(j)}}{\leq} nH(Y_1|Z)+I(M_2;Z^n|M_1)+n(\epsilon_n+\tau_n)&\\
&\leq nH(Y_1|Z)+H(M_2|M_1)+n(\epsilon_n+\tau_n)&\\
&=nH(Y_1|Z)+nR_2+n(\epsilon_n+\tau_n)&
\end{flalign*}
where (g) follows from the Markov chain $(M_1,M_2,Y_{1}^{i+1},Z^{i-1},Z_{i+1}^{n}) \rightarrow X_i \rightarrow (Y_{1i},Z_i)$, (h) follows from the time-sharing argument, (i) follows from the fact that $H(Y_1|X,Z)=0$ since $Y_1$ is a deterministic function of $X$ and (j) follows from (\ref{aa5}).

On the other hand, we have
\begin{flalign*}
nR_1&=H(M_1)&\\
&=I(M_1;Y_2^n)+H(M_1|Y_2^n)&\\
&\overset{\text{(k)}}{\leq}I(M_1;Y_2^n)+2n\epsilon_n&\\
&=I(M_1;Y_2^n)+I(M_1;Z^n)-I(M_1;Z^n)+2n\epsilon_n&\\
&\overset{\text{(l)}}{\leq}I(M_1;Y_2^n)-I(M_1;Z^n)+n(2\epsilon_n+\tau_n)&\\
&\leq I(M_1;Y_2^n,Z^n)-I(M_1;Z^n)+n(2\epsilon_n+\tau_n)&\\
&= I(M_1;Y_2^n|Z^n)+n(2\epsilon_n+\tau_n)&\\
&=\sum_{i=1}^{n}I(M_1;Y_{2i}|Y_2^{i-1},Z^n)+n(2\epsilon_n+\tau_n)&\\
&\leq\sum_{i=1}^{n}I(M_1,Y_2^{i-1},Z^{i-1},Z_{i+1}^n;Y_{2i}|Z_i)+n(2\epsilon_n+\tau_n)&\\
&\overset{\text{(m)}}{\leq} \sum_{i=1}^{n}I(X_i;Y_{2i}|Z_i)+n(2\epsilon_n+\tau_n)&\\
&\overset{\text{(n)}}{\leq} nI(X;Y_{2}|Z)+n(2\epsilon_n+\tau_n)&\\
&\overset{\text{(o)}}{=} nH(Y_2|Z)+n(2\epsilon_n+\tau_n)&
\end{flalign*}
where (k) follows from (\ref{aa4}), (l) follows (\ref{aa5}), (m) follows from the Markov chain $(M_1,Y_2^{i-1},Z^{i-1},Z_{i+1}^n) \rightarrow X_i \rightarrow (Y_{2i},Z_i)$, (n) follows from the time-sharing argument and (o) follows from the fact that $H(Y_2|X)=0$ since $Y_2$ is a deterministic function of $X$.

Next, we have
\begin{flalign*}
nR_2&=H(M_2)&\\
&=I(M_2;Y_2^n)+H(M_2|Y_2^n)&\\
&\overset{\text{(p)}}{\leq}I(M_2;Y_2^n)+n\epsilon_n&\\
&=I(M_2;Y_2^n)-I(M_2;Z^n)+I(M_2;Z^n)+n\epsilon_n&\\
&\overset{\text{(q)}}{\leq}I(M_2;Y_2^n)-I(M_2;Z^n)+n(\epsilon_n+\tau_n)&\\
&\leq I(M_2;Y_2^n,Z^n)-I(M_2;Z^n)+n(\epsilon_n+\tau_n)&\\
&= I(M_2;Y_2^n|Z^n)+n(\epsilon_n+\tau_n)&\\
&=\sum_{i=1}^{n}I(M_2;Y_{2i}|Y_2^{i-1},Z^n)+n(\epsilon_n+\tau_n)&\\
&\leq \sum_{i=1}^{n}I(M_2,Y_2^{i-1},Z^{i-1},Z_{i+1}^{n};Y_{2i}|Z_i)+n(\epsilon_n+\tau_n)&\\
&\overset{\text{(r)}}{\leq} \sum_{i=1}^{n}I(X_i;Y_{2i}|Z_i)+n(\epsilon_n+\tau_n)&\\
&\overset{\text{(s)}}{\leq} nI(X;Y_{2}|Z)+n(\epsilon_n+\tau_n)&\\
&\overset{\text{(t)}}{=} nH(Y_2|Z)+n(\epsilon_n+\tau_n)&
\end{flalign*}
where (p) follows from (\ref{aa2}), (q) follows from (\ref{aa6}), (r) follows from the Markov chain $(M_2,Y_2^{i-1},Z^{i-1},Z_{i+1}^{n}) \rightarrow X_i \rightarrow (Y_{2i},Z_{i})$, (s) follows from the time-sharing argument and (t) follows from the fact that $H(Y_2|X)=0$ since $Y_2$ is a deterministic function of $X$.

Lastly, for the sum rate bound, we have
\begin{flalign*}
n(R_1+R_2)&=H(M_1,M_2)&\\
&=H(M_2)+H(M_1|M_2)&\\
&=I(M_2;Y_2^n)+H(M_2|Y_2^n)+I(M_1;Y_2^n|M_2)&\\&\quad+H(M_1|Y_2^n,M_2)&\\
&\overset{\text{(u)}}{\leq}I(M_2;Y_2^n)+I(M_1;Y_2^n|M_2)+2n\epsilon_n&\\
&=I(M_1,M_2;Y_2^n)+2n\epsilon_n&\\
&=\sum_{i=1}^{n}I(M_1,M_2;Y_{2i}|Y_{2}^{i-1})+2n\epsilon_n&\\
&\leq \sum_{i=1}^{n}I(M_1,M_2,Y_{2}^{i-1};Y_{2i})+2n\epsilon_n&\\
&\overset{\text{(v)}}{\leq} \sum_{i=1}^{n}I(X_i;Y_{2i})+2n\epsilon_n&\\
&\overset{\text{(w)}}{\leq} nI(X;Y_{2})+2n\epsilon_n&\\
&\overset{\text{(x)}}{=} nH(Y_{2})+2n\epsilon_n&
\end{flalign*}
where (u) follows from (\ref{aa2}) and (\ref{aa3}), (v) follows from the Markov chain $(M_1,M_2,Y_{2}^{i-1}) \rightarrow X_i \rightarrow Y_{2i}$, (w) follows from the time-sharing argument and (x) follows from the fact that $H(Y_2|X)=0$ since $Y_2$ is a deterministic function of $X$.
\end{IEEEproof}

\section{}\label{AppC}
In this section, we will present the converse proof of Theorem~\ref{theorem3}.
\begin{IEEEproof}[Proof of Theorem~\ref{theorem3}] \textit{(Converse)}
	By Fano's inequality \cite[p. 19]{cref6}, we have the reliability constraints
	(\ref{aa1} and (\ref{aa2}).
	From (\ref{aa2}) and the fact that $Y_2$ is a degraded version of $Y_1$, we have
	\begin{flalign} 
	H(M_2|Y_1^n)\leq H(M_2|Y_2^n)\leq n\epsilon_n
	\end{flalign}
	and consequently
	\begin{flalign} 
	H(M_2|Y_1^n)\leq n\epsilon_n.\label{bb3}
	\end{flalign}
	With this, we establish 
	\begin{flalign}
	H(M_1|Y_1^n)&\leq H(M_1|Y_1^n)+H(M_2|Y_1^n,M_1)\nonumber\\
	&=H(M_1,M_2|Y_1^n)\nonumber\\
	&=H(M_2|Y_1^n)+H(M_1|Y_1^n,M_2)\nonumber\\
	&\overset{\text{(a)}}{\leq}H(M_2|Y_1^n)+n\epsilon_n\nonumber\\
	&\overset{\text{(b)}}{\leq}2n\epsilon_n,\label{bb4}
	\end{flalign}
	where (a) follows from (\ref{aa1}) and (b) follows from (\ref{bb3}). At the same time, due to individual secrecy, we have the secrecy constraints 
	\begin{flalign}
	I(M_1;Z^n)\leq n\tau_n\label{bb5},\\
	I(M_2;Z^n)\leq n\tau_n\label{bb6}.
	\end{flalign}
	
	Using (\ref{aa1}), (\ref{aa2}) and (\ref{bb3})--(\ref{bb6}), we establish the following.
	\begin{flalign*}
	nR_1&=H(M_1)&\\
	&=I(M_1;Y_1^n)+H(M_1|Y_1^n)&\\
	&\overset{\text{(c)}}{\leq}I(M_1;Y_1^n)+2n\epsilon_n&\\
	&=I(M_1;Y_1^n)+I(M_1;Z^n)-I(M_1;Z^n)+2n\epsilon_n&\\
	&\overset{\text{(d)}}{\leq}I(M_1;Y_1^n)-I(M_1;Z^n)+n(2\epsilon_n+\tau_n)&\\
	&\leq I(M_1;Y_1^n,Z^n)-I(M_1;Z^n)+n(2\epsilon_n+\tau_n)&\\
	&= I(M_1;Y_1^n|Z^n)+n(2\epsilon_n+\tau_n)&\\
	&=\sum_{i=1}^{n}I(M_1;Y_{1i}|Y_1^{i-1},Z^n)+n(2\epsilon_n+\tau_n)&\\
	&\leq\sum_{i=1}^{n}I(M_1,Y_1^{i-1},Z^{i-1},Z_{i+1}^n;Y_{1i}|Z_i)+n(2\epsilon_n+\tau_n)&\\
	&\overset{\text{(e)}}{\leq} \sum_{i=1}^{n}I(X_i;Y_{1i}|Z_i)+n(2\epsilon_n+\tau_n)&\\
	&\overset{\text{(f)}}{\leq} nI(X;Y_{1}|Z)+n(2\epsilon_n+\tau_n)&\\
	&\overset{\text{(g)}}{=} nH(Y_1|Z)+n(2\epsilon_n+\tau_n)&
	\end{flalign*}
	where (c) follows from (\ref{bb4}), (d) follows (\ref{bb5}), (e) follows from the Markov chain $(M_1,Y_1^{i-1},Z^{i-1},Z_{i+1}^n) \rightarrow X_i \rightarrow (Y_{1i},Z_i)$, (f) follows from the time-sharing argument and (g) follows from the fact that $H(Y_1|X)=0$ since $Y_1$ is a deterministic function of $X$.
	
	Next, we have
	\begin{flalign*}
	nR_2&=H(M_2)&\\
	&=I(M_2;Y_2^n)+H(M_2|Y_2^n)&\\
	&\overset{\text{(h)}}{\leq}I(M_2;Y_2^n)+n\epsilon_n&\\
	&=I(M_2;Y_2^n)-I(M_2;Z^n)+I(M_2;Z^n)+n\epsilon_n&\\
	&\overset{\text{(i)}}{\leq}I(M_2;Y_2^n)-I(M_2;Z^n)+n(\epsilon_n+\tau_n)&\\
	&\leq I(M_2;Y_2^n,Z^n)-I(M_2;Z^n)+n(\epsilon_n+\tau_n)&\\
	&= I(M_2;Y_2^n|Z^n)+n(\epsilon_n+\tau_n)&\\
	&=\sum_{i=1}^{n}I(M_2;Y_{2i}|Y_2^{i-1},Z^n)+n(\epsilon_n+\tau_n)&\\
	&\leq \sum_{i=1}^{n}I(M_2,Y_2^{i-1},Z^{i-1},Z_{i+1}^{n};Y_{2i}|Z_i)+n(\epsilon_n+\tau_n)&\\
	&\overset{\text{(j)}}{\leq} \sum_{i=1}^{n}I(X_i;Y_{2i}|Z_i)+n(\epsilon_n+\tau_n)&\\
	&\overset{\text{(k)}}{\leq} nI(X;Y_{2}|Z)+n(\epsilon_n+\tau_n)&\\
	&\overset{\text{(l)}}{=} nH(Y_2|Z)+n(\epsilon_n+\tau_n)&
	\end{flalign*}
	where (h) follows from (\ref{aa2}), (i) follows from (\ref{bb6}), (j) follows from the Markov chain $(M_2,Y_2^{i-1},Z^{i-1},Z_{i+1}^{n}) \rightarrow X_i \rightarrow (Y_{2i},Z_{i})$, (k) follows from the time-sharing argument and (l) follows from the fact that $H(Y_2|X)=0$ since $Y_2$ is a deterministic function of $X$.
	
	Lastly, for the sum rate bound, we have
	\begin{flalign*}
	n(R_1+R_2)&=H(M_1,M_2)&\\
	&=H(M_2)+H(M_1|M_2)&\\
	&=I(M_2;Y_1^n)+H(M_2|Y_1^n)+I(M_1;Y_1^n|M_2)&\\&\quad+H(M_1|Y_1^n,M_2)&\\
	&\overset{\text{(m)}}{\leq}I(M_2;Y_1^n)+I(M_1;Y_1^n|M_2)+2n\epsilon_n&\\
	&=I(M_1,M_2;Y_1^n)+2n\epsilon_n&\\
	&=\sum_{i=1}^{n}I(M_1,M_2;Y_{1i}|Y_{1}^{i-1})+2n\epsilon_n&\\
	&\leq \sum_{i=1}^{n}I(M_1,M_2,Y_{1}^{i-1};Y_{1i})+2n\epsilon_n&\\
	&\overset{\text{(n)}}{\leq} \sum_{i=1}^{n}I(X_i;Y_{1i})+2n\epsilon_n&\\
	&\overset{\text{(o)}}{\leq} nI(X;Y_{1})+2n\epsilon_n&\\
	&\overset{\text{(p)}}{=} nH(Y_{1})+2n\epsilon_n&
	\end{flalign*}
	where (m) follows from (\ref{aa1}) and (\ref{bb3}), (n) follows from the Markov chain $(M_1,M_2,Y_{1}^{i-1}) \rightarrow X_i \rightarrow Y_{1i}$, (o) follows from the time-sharing argument and (p) follows from the fact that $H(Y_1|X)=0$ since $Y_1$ is a deterministic function of $X$.
\end{IEEEproof}

\enlargethispage{-12cm} 

\bibliographystyle{IEEEtran}
\bibliography{citations}

\begin{thebibliography}{10}
\providecommand{\url}[1]{#1}
\csname url@samestyle\endcsname
\providecommand{\newblock}{\relax}
\providecommand{\bibinfo}[2]{#2}
\providecommand{\BIBentrySTDinterwordspacing}{\spaceskip=0pt\relax}
\providecommand{\BIBentryALTinterwordstretchfactor}{4}
\providecommand{\BIBentryALTinterwordspacing}{\spaceskip=\fontdimen2\font plus
\BIBentryALTinterwordstretchfactor\fontdimen3\font minus
  \fontdimen4\font\relax}
\providecommand{\BIBforeignlanguage}[2]{{%
\expandafter\ifx\csname l@#1\endcsname\relax
\typeout{** WARNING: IEEEtran.bst: No hyphenation pattern has been}%
\typeout{** loaded for the language `#1'. Using the pattern for}%
\typeout{** the default language instead.}%
\else
\language=\csname l@#1\endcsname
\fi
#2}}
\providecommand{\BIBdecl}{\relax}
\BIBdecl

\bibitem{iref1}
I.~Csiszar and J.~Korner, ``Broadcast channels with confidential messages,''
  \emph{IEEE Transactions on Information Theory}, vol.~24, no.~3, pp. 339--348,
  May 1978.

\bibitem{iref2}
Y.~K. Chia and A.~E. Gamal, ``3-receiver broadcast channels with common and
  confidential messages,'' in \emph{2009 IEEE International Symposium on
  Information Theory}, June 2009, pp. 1849--1853.

\bibitem{Schaefer_Boche14}
R.~F. Schaefer and H.~Boche, ``Robust broadcasting of common and confidential
  messages over compound channels: Strong secrecy and decoding performance,''
  \emph{IEEE Transactions on Information Forensics and Security}, vol.~9,
  no.~10, pp. 1720--1732, Oct 2014.

\bibitem{cref1}
Y.~Chen, O.~O. Koyluoglu, and A.~Sezgin, ``Individual secrecy for the broadcast
  channel,'' \emph{IEEE Transactions on Information Theory}, vol.~63, no.~9,
  pp. 5981--5999, Sept 2017.

\bibitem{iref13}
R.~F. Wyrembelski, A.~Sezgin, and H.~Boche, ``Secrecy in broadcast channels
  with receiver side information,'' in \emph{2011 Conference Record of the
  Forty Fifth Asilomar Conference on Signals, Systems and Computers
  (ASILOMAR)}, Nov 2011, pp. 290--294.

\bibitem{iref14}
A.~S. Mansour, R.~F. Schaefer, and H.~Boche, ``Joint and individual secrecy in
  broadcast channels with receiver side information,'' in \emph{2014 IEEE 15th
  International Workshop on Signal Processing Advances in Wireless
  Communications (SPAWC)}, June 2014, pp. 369--373.

\bibitem{cref2}
Y.~Chen, O.~O. Koyluoglu, and A.~Sezgin, ``Individual secrecy for broadcast
  channels with receiver side information,'' \emph{IEEE Transactions on
  Information Theory}, vol.~63, no.~7, pp. 4687--4708, July 2017.

\bibitem{iref15}
A.~S. Mansour, R.~F. Schaefer, and H.~Boche, ``Secrecy measures for broadcast
  channels with receiver side information: Joint vs individual,'' in \emph{2014
  IEEE Information Theory Workshop (ITW 2014)}, Nov 2014, pp. 426--430.

\bibitem{Me1}
\BIBentryALTinterwordspacing
J.~Y. Tan, L.~Ong, and B.~Asadi, ``A simplified coding scheme for the broadcast
  channel with complementary receiver side information under individual secrecy
  constraints,'' 2018. [Online]. Available:
  \url{http://arxiv.org/abs/1801.04033}
\BIBentrySTDinterwordspacing

\bibitem{cref6}
A.~E. Gamal and Y.-H. Kim, \emph{Network Information Theory}.\hskip 1em plus
  0.5em minus 0.4em\relax New York, NY, USA: Cambridge University Press, 2012.

\bibitem{cref5}
K.~Marton, ``A coding theorem for the discrete memoryless broadcast channel,''
  \emph{IEEE Transactions on Information Theory}, vol.~25, no.~3, pp. 306--311,
  May 1979.

\bibitem{cref8}
C.~E. Shannon, ``Communication theory of secrecy systems,'' \emph{The Bell
  System Technical Journal}, vol.~28, no.~4, pp. 656--715, Oct 1949.

\bibitem{cref3}
A.~Carleial and M.~Hellman, ``A note on wyner's wiretap channel (corresp.),''
  \emph{IEEE Transactions on Information Theory}, vol.~23, no.~3, pp. 387--390,
  May 1977.

\bibitem{cref4}
A.~D. Wyner, ``The wire-tap channel,'' \emph{The Bell System Technical
  Journal}, vol.~54, no.~8, pp. 1355--1387, Oct 1975.

\bibitem{Mansour_Schaefer_Boche16}
A.~S. Mansour, R.~F. Schaefer, and H.~Boche, ``On the individual secrecy
  capacity regions of the general, degraded, and gaussian multi-receiver
  wiretap broadcast channel,'' \emph{IEEE Transactions on Information Forensics
  and Security}, vol.~11, no.~9, pp. 2107--2122, Sept 2016.

\bibitem{Liu_Maric_Spasojevic_Yates08}
R.~Liu, I.~Maric, P.~Spasojevic, and R.~D. Yates, ``Discrete memoryless
  interference and broadcast channels with confidential messages: Secrecy rate
  regions,'' \emph{IEEE Transactions on Information Theory}, vol.~54, no.~6,
  pp. 2493--2507, June 2008.

\bibitem{Kramer08}
\BIBentryALTinterwordspacing
G.~Kramer, ``Topics in multi-user information theory,'' \emph{Found. Trends
  Commun. Inf. Theory}, vol.~4, no. 4-5, pp. 265--444, Apr. 2008. [Online].
  Available: \url{http://dx.doi.org/10.1561/0100000028}
\BIBentrySTDinterwordspacing

\end{thebibliography}

	







\end{document}